\documentclass[a4paper,11pt]{article}

\usepackage{jheppub}
\usepackage{amsmath}
\usepackage{amsfonts}
\usepackage{amssymb}

\usepackage{graphicx}
\usepackage{dcolumn}
\usepackage{bm}
\usepackage{epsfig}

\newcommand{\dhd}{{\textstyle d}
	\lower.03ex\hbox{\kern-0.38em$^{\scriptstyle-}$}\kern-0.05em{}}
\newcommand{\dbar}{{\textstyle \delta}
	\lower.03ex\hbox{\kern-0.38em$^{\scriptstyle-}$}\kern-0.05em{}}
\newcommand{\half}{{1\over 2}}

\newcommand{\baru}{{\bar u}}

\newcommand{\bamma}{{\bar \gamma}}

\newcommand{\calf}{{\cal F}}

\newcommand{\calm}{{\cal M}}
\newcommand{\caln}{{\cal N}} 
\newcommand{\calo}{{\cal O}}

\newcommand{\cals}{{\cal S}} 
 
\newcommand{\calu}{{\cal U}} 
\newcommand{\calv}{{\cal V}}

\newcommand{\barpsi}{{\bar \psi}}

\newcommand{\hatx}{{\hat x}} 
\newcommand{\haty}{{\hat y}} 
\newcommand{\hatz}{{\hat z}} 
 
\newcommand{\hatX}{{\hat X}} 
\newcommand{\hatY}{{\hat Y}}

\newcommand{\tildeu}{{\tilde u}}

\newcommand \ket [1] {|{#1}\rangle}
\newcommand \bra [1] {\langle {#1}|}

\newcommand \sslash [1] {\slash\hspace{-0.2cm}{#1}}

\newcommand{\ssx}{\sslash{x}}
\newcommand{\ssy}{\sslash{y}}
\newcommand{\ssz}{\sslash{z}}

\newcommand{\ssxi}{\sslash{\xi}}

\newcommand{\Tr}{{\rm Tr}} 
\newcommand{\tr}{{\rm tr}}

\begin{document}
	
	\title{Pseudo and quasi quark PDF in the BFKL approximation}

\author[]{Giovanni Antonio Chirilli}
\affiliation[]{Departamento de F\'{\i}sica de Part\'{\i}culas and IGFAE,\\
Universidade de Santiago de Compostela,\\
 15782 Santiago de Compostela, Galicia, Spain}
\emailAdd{giovanniantonio.chirilli@usc.es}

\abstract{

I examine the high-energy behavior of the Ioffe-time distribution for the quark bi-local space-like separated operator using the high-energy operator product expansion. These findings have significant implications for lattice calculations, which require extrapolation for large Ioffe-time values. 
I perform an explicit Fourier transform for both the pseudo-PDF and quasi-PDF, and investigate their behavior within the first two leading twist contributions.

I show that the quark pseudo-PDF captures the BFKL resummation (resummation of all twists) and exhibits a rising behavior for small $x_B$ values, while the quasi-PDF presents a different behavior. I demonstrate that an appropriate small-$x_B$ behavior cannot be achieved solely through DGLAP dynamics, emphasizing the importance of all-twist resummation. This study provides valuable insights into quark non-local operators' high-energy behavior and the limitations of lattice calculations in this context.
 
}

\maketitle
	
\section{Introduction}

Over the past decade, the investigation of Euclidean-separated, gauge-invariant, bi-local operators via lattice gauge formalism has garnered significant interest due to its ability to provide direct access to parton distribution functions (PDFs) from first principles
(for a review see~\cite{Cichy:2018mum, Ji:2020ect, Cichy:2021lih, Constantinou:2022yye}). The foundation for this surge in activity lies in the observation that space-like separated operators can be examined through lattice QCD formalism, and that in the infinite momentum frame, they reduce to the conventional light-cone operators through which PDFs are defined~\cite{Ji:2013dva}. Deviations from the infinite momentum frame emerge as inverse powers of the large parameter of the boost, suggesting that such corrections can be systematically suppressed.

The distributions introduced in~\cite{Ji:2013dva}, known as quasi-PDFs, were later complemented by alternative PDFs called pseudo-PDFs in~\cite{Radyushkin:2017cyf}. The Bjorken-$x$ ($x_B$) dependence of these two distributions has been extensively studied in recent years~\cite{Chen:2016utp, Orginos:2017kos, Joo:2020spy, Ji:2017oey, HadStruc:2021wmh}; however, lattice formalism is unlikely to provide access to their behavior at small $x_B$ values. 

We are going to demonstrate (preliminary result have been presented in~\cite{Chirilli:2022dzt}) that despite being defined through the same 
space-like separated bi-local operators~\cite{Braun:2007wv, Bali:2017gfr, Bali:2018spj, DelDebbio:2020rgv}, 
quark quasi- and pseudo-PDFs exhibit markedly different behavior at small $x_B$ values. 

The importance of understanding the small-$x_B$ behavior of quark pseudo- and quasi-PDFs cannot be overstated, 
even when the corresponding behavior for gluon pseudo- and quasi-PDFs has been established~\cite{Chirilli:2021euj}. 
This is primarily due to the fact that most lattice calculations focus on quark distributions rather than gluon ones. 
Comprehensive knowledge of small-$x_B$ behavior for quark distributions is essential for providing a more complete picture of the underlying dynamics in QCD, ultimately leading to more accurate predictions and a deeper understanding of the fundamental interactions within the strong force. Moreover, insights into the quark pseudo- and quasi-PDFs' behavior at small-$x_B$ values will be indispensable for the interpretation of experimental results from future QCD colliders probing this regime, such as the Electron-Ion Collider (EIC) in the USA~\cite{Accardi:2012qut}, the EIC in China~\cite{Anderle:2021wcy}, 
and the Large Hadron electron Collider (LHeC) in Europe~\cite{LHeC:2020van}. 

The small-$x_B$ behavior of deep inelastic scattering (DIS) structure functions can be calculated using the high-energy operator product expansion (OPE)~\cite{Balitsky:1995ub}, wherein the T-product of two electromagnetic currents is expanded in terms of coefficient functions, also known as impact factors, convoluted with the matrix elements of infinite Wilson lines. The evolution equation of the matrix elements of infinite Wilson lines with respect to the rapidity factorization parameter is the 
BK equation~\cite{Balitsky:1995ub, Kovchegov:1999yj, Jalilian-Marian:1997gr, Ferreiro:2001qy, Iancu:2000hn}, 
which addresses small-$x_B$ leading-log resummation through its linear term and the unitarity property of the theory through its non-linear term. Due to its non-linear nature, the BK equation is a generalization of the BFKL equation.

The high-energy OPE is a complementary procedure to the OPE in non-local (finite gauge-link) operators adopted in the Bjorken limit, where the factorization scale is the transverse momentum of the fields and the evolution equation of the finite gauge-link with respect to the factorization parameter is the DGLAP evolution equation.

In the DGLAP regime, the structure functions of DIS are governed by the anomalous dimensions of twist-2 operators, while in the BFKL regime, the structure functions receive contributions from an infinite series of twist expansion. By moving beyond the BFKL limit to the overlapping region, one can obtain the anomalous dimension of twist-2 (and potentially higher twist) operators in the small-$x_B$ limit from the small-$x_B$ structure function. With the analytic continuation of the twist-2 local operator to non-integer spin, the local operators become non-local light-ray operators, making the prediction of the small-$x_B$ limit of the twist-2 anomalous dimension from the BFKL resummation more transparent~\cite{Balitsky:2013npa}. In this work, we demonstrate that this formalism enables us to obtain the leading (LT) 
and next-to-leading twist (NLT) corrections from the full BFKL resummed results for the pseudo- and quasi-PDFs.

In the Appendix we will study correlation function of two light-ray 
operators~\cite{Balitsky:1987bk, Balitsky:2014sqe, Balitsky:2013npa, Balitsky:2015tca, Balitsky:2015oux, Balitsky:2018irv, Kravchuk:2018htv}, in the BFKL limit.
Investigating this correlation function in this limit is of great importance, particularly as recent works have primarily focused on the gluon case. Studying the quark correlation function provides a more comprehensive understanding of the underlying physics in the BFKL regime and is crucial for elucidating the behavior of four-point functions in the general case of conformal field theories.

The organization of the paper is as follows. In Section \ref{sec:Ioffetime}, we provide a concise introduction to the high-energy OPE formalism and demonstrate its application to the Ioffe-time distribution.
In Section \ref{sec:Ioffetwist}, we examine the Ioffe-time distribution in the context of leading and next-to-leading twist approximations.
In order to reinforce our conclusions pertaining to the high-energy behavior of the Ioffe-time amplitude reached using the
Golec-Biernat Wusthoff model, in section \ref{sec:IoffetimePhoton}, our focus will shift to the photon impact factor model, which distinctly, does not embody saturation mechanisms.
Sections \ref{sec:quarkpseudo} and \ref{sec:quarkquasi} are dedicated to the investigation of pseudo and quasi PDFs, respectively, within the BFKL limit and in the leading and next-to-leading twist approximations.
Finally, in Section \ref{sec:conclusions}, we present our conclusions and discuss the implications of our findings.

\section{Ioffe-time distribution in the BFKL limit}
\label{sec:Ioffetime}

The quark distribution is defined through the light-cone matrix element~\cite{Balitsky:1987bk}:
\begin{eqnarray}
&&\hspace{-5mm}
\langle P|\bar\psi(z)\ssz[z,0]\psi(0)|P\rangle
=2P\!\cdot\! z\!\!\int_{-1}^1\!\! dx_B\,e^{iP\cdot z\, x_B}Q(x_B)
\label{def-qpdf2}
\end{eqnarray}
with $z^2=0$, $P$ the hadronic momentum, and with the gauge link defined as
\begin{eqnarray}
\hspace{-0.3cm}
[x,y] = {\rm P}{\rm exp}\left\{ig\!\!\int_0^1\!\! du\,(x-y)^\mu A_\mu \left(x+(1-u)y\right)\right\}\,.
\end{eqnarray} 
The distribution $Q(x_B)$ is defined through the quark distribution, $D_q$, and anti-quark distribution, $D_{\bar{q}}$ as
\begin{eqnarray}
Q(x_B)~=~\theta(x_B)D_q(x_B)-\theta(-x_B)D_{\bar{q}}(-x_B)\,.
\end{eqnarray}

In order to calculate the low-$x_B$ behavior of the Deep Inelastic Scattering (DIS) structure function, one can employ the high-energy Operator Product Expansion (OPE) framework. In this approach, the T-product of two electromagnetic currents, 
${\rm T}{J^\mu(x)J^\nu(y)}$, is expanded in terms of coefficient functions (impact factors), convoluted
with the matrix elements of infinite Wilson lines, which are the relevant operators in the high-energy (Regge) limit. 
The evolution equations governing the matrix elements of infinite Wilson lines with respect to the factorization parameter, or rapidity, 
are the BK-JIMWLK evolution equations~\cite{Balitsky:1995ub, Kovchegov:1999yj, Jalilian-Marian:1997gr, Ferreiro:2001qy, Iancu:2000hn}. 
This equation accounts for both leading-log resummation and preservation 
of the unitarity property of the theory. The BK equation is considered a generalization of the BFKL equation, which, 
although it handles leading-log resummation, results in a violation of unitarity.

The high-energy OPE is analogous to the OPE in non-local (finite gauge-link) operators employed in the Bjorken limit. In this case, the factorization scale is in transverse momentum, and the evolution equation of the finite gauge-link with respect to the factorization parameter is the DGLAP evolution equation.

To examine the high-energy behavior of the Ioffe-time distribution and, consequently, the low-$x_B$ behavior of pseudo- and quasi-PDFs, we consider Lorentz decomposition of the structure function defined through the bi-local operator~\cite{Joo:2020spy}:
\begin{eqnarray}
\bra{P}\barpsi(z)\gamma^\mu[z, 0]\psi(0)\ket{P} = 2P^\mu \calm(\varrho,z^2) +2z^\mu\caln(\varrho,z^2)\,.
\label{lorentzdecomp}
\end{eqnarray}
Here, $\varrho = z\cdot P$, and $P$ represents the hadronic momentum.  

In the high-energy limit, the $0$ component and the $3$ component are indistinguishable. 
Let us introduce the light-cone vectors $n$ and $n'$ such that $n^2=n'^2=0$, $n\cdot n'=1$, and $x\cdot n'=x^+$, $x^-\cdot n=x^-$. Here, $x^\pm={x^0\pm x^3\over \sqrt{2}}$. In this way, a generic vector can be written as $x^\mu=x^+n^\mu + x^- n'^\mu + x^\mu_\perp$, 
where the transverse coordinate is $x^\mu_\perp = (0,x^1,x^2,0)$.
We adopt the high-energy limit in which the largest light-cone component of the target's momentum is $P^-$.
Under this large boost, we have 
$\bar{\psi}(x^+,x^-,x_\perp)\gamma^\mu\psi(x^+,x^-,x_\perp)\to \bar{\psi}(x^+,x_\perp)n'^\mu\gamma^-\psi(x^+,x_\perp)$.
Projection of (\ref{lorentzdecomp}) along $z^\mu$ gives
\begin{eqnarray}
{z_\mu\over 2\varrho}\bra{P}\barpsi(z)\gamma^\mu[z, 0]\psi(0)\ket{P} = \calm(\varrho,z^2) + {z^2\over \varrho}\caln(\varrho,z^2)
\label{defoperator}\,.
\end{eqnarray}
In the high-energy limit ${z^2\over \varrho}\to {2z^+z^--z^2_\perp\over z^+P^-} \to 0$. Consequently, 
the component of the structure function (\ref{lorentzdecomp}) that survives after the boost is $\calm(\varrho,z^2)$
which contains the leading twist as well as higher twist. On the other hand, the component $\caln(\varrho, z^2)$ contains only
higher twists which are suppressed in the high-energy limit. 

At high-energy, the bi-local operator is written in terms of integration along the longitudinal direction as
\begin{eqnarray}
&&{1\over 2P^-}\bra{P}\barpsi(L,x_\perp)\gamma^-[Ln^\mu+x_\perp, 0]\psi(0)\ket{P + \epsilon^- n'}\,2\pi\delta(\epsilon^-)
\nonumber\\
&& = \int_0^{+\infty} \!\!dx^+\!\! \int_{-\infty}^0\!\!dy^+\delta(x^+-y^+-L)
\nonumber\\
&&\times{1\over 2P^-}\bra{P}\barpsi(x^+,x_\perp)\gamma^-[x^+n^\mu+x_\perp, y^+n^\mu+0_\perp]\psi(y^+,0_\perp)\ket{P+\epsilon^-n'}
\label{IoffeAmpOperator}
\end{eqnarray}
where, as discussed above,  at high-energy we can write ${\ssz\over 2z\cdot P}\to {\gamma^-\over 2P^-}$.
We can now apply the high-energy OPE to (\ref{IoffeAmpOperator}).
This method was also used in~\cite{Chirilli:2021euj} for the study of
the high-energy behavior of the gluon Ioffe-time distribution. 

To begin calculating the impact factor diagram, we consider the operator in equation (\ref{IoffeAmpOperator}) in the external gluon field, which, at high-energy, shrinks to a shock-wave (represented by the red vertical band in the figure). In the high-energy (Regge) limit, the main degree of freedom are gluons,
so it is natural to assume that the field of the target state is predominantly made of gluons~.
For this purpose, we need the quark propagator in the shock-wave in coordinate space~\cite{Balitsky:1995ub}
\begin{eqnarray}
\langle\psi(x)\barpsi(y)\rangle \stackrel{x^+>0>y^+}{=} 
\int d^4z\delta(z^+){\ssx-\ssz\over 2\pi^2[(x-z)^2-i\epsilon]^2}
\sslash{n}'U_z{\ssy-\ssz\over 2\pi^2[(y-z)^2-i\epsilon]^2}\,.
\label{LO-DIS-OPE}
\end{eqnarray}
\begin{figure}[t!]
	\begin{center}
		\includegraphics[width=2.0in]{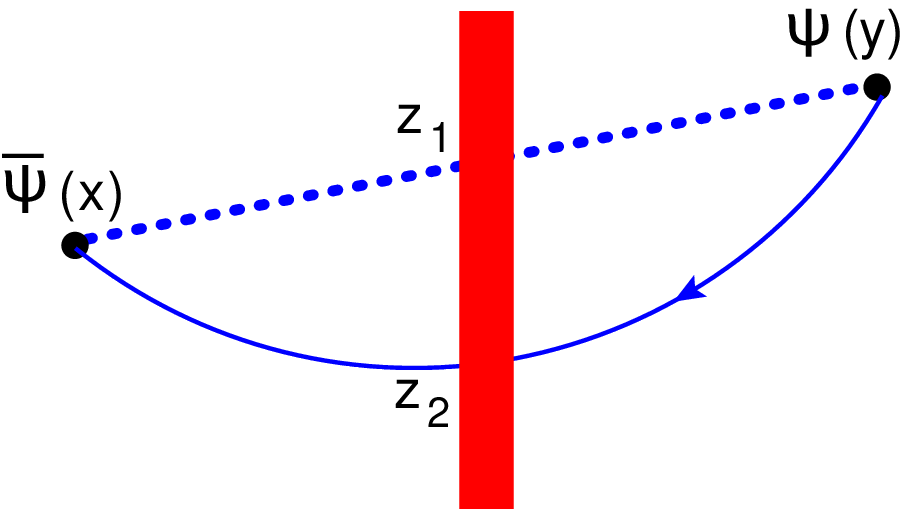}
		\caption{Diagram for the LO impact factor.
			We indicate in blue the quantum fields and in red the classical background field which, in the leading accuracy, is made by gluons.}
		\label{Fig:quarkLOif}
	\end{center}
\end{figure}
where $U_x = U(x_\perp) = {\rm P}{\rm exp}\left\{ig\int_{-\infty}^{+\infty} dx^+A^-(x^+,x_\perp)\right\}$.

Functionally integrating over the quantum field, which are the quark fields at point $x$ and point $y$,
the Ioffe-time non-local operator, eq. (\ref{IoffeAmpOperator}), becomes
\begin{eqnarray}
&&\hspace{-2cm}\int dx^+ dy^+\delta(x^+-y^+-L)
\langle\barpsi(x^+,x_\perp)\gamma^-[x^+n^\mu+x_\perp, y^+n^\mu+y_\perp]\psi(y^+,y_\perp)\rangle_A
\nonumber\\
&&\hspace{-2cm}
\stackrel{y^+<0<x^+}{=} \int_0^{+\infty} {dx^+\over {x^+}^2}
\int_{-\infty}^0{dy^+\over {y^+}^2}\,\delta(x^+-y^+-L)
\nonumber\\
&&\hspace{-1cm}
\times\!\int\! {d^2z_2\over 2\pi^3}
{-i\,\tr\big\{\gamma^-(\haty-\hatz_2)\gamma^+ (\hatx-\haty_2)\}\over \Big[{(y-z_2)^2_\perp\over |y^+|} + {(x-z_2)^2_\perp\over x^+}+i\epsilon\Big]^3}\,
\langle\Tr\{U_{z_1}U^\dagger_{z_2}\}\rangle_A
\label{1ststepOPE}
\end{eqnarray}	
The subscript $A$ on the angle bracket means that the operator is being evaluated in the background of the gluon field.
In eq. (\ref{1ststepOPE}) we have reduced (expanded) the operator on the left-hand-side (LHS) of the equal sign into 
a convolution between a coefficient (the impact factor) and new operators given by the trace in the fundamental representation of two
infinite Wilson lines, one at the transverse point $z_{1\perp}$ and another one at the transverse point $z_{2\perp}$.
Let us define
$X_{1\perp} = x_\perp-z_{1\perp}$ and $Y_{1\perp}=y_\perp-z_{1\perp}$, and from eq. (\ref{1ststepOPE}) we arrive at
\begin{eqnarray}
&&\hspace{-1cm}\int dx^+ dy^+\delta(x^+-y^+-L)
\langle\barpsi(x^+,x_\perp)\gamma^-[x^+n^\mu+x_\perp, y^+n^\mu+y_\perp]\psi(y^+,y_\perp)\rangle_A
\label{quark-startinpoint}
\\
&&\hspace{-1cm}
=  \int_0^{+\infty} {dx^+\over {x^+}^2} \int_{-\infty}^0d{y^+\over {y^+}^2}\, \delta(x^+-y^+-L)
\int {d^2z_2\over 2\pi^3}
{- 4\,i\,(X_2,Y_2)_\perp\over \Big[{(y-z_2)^2_\perp\over |y^+|} + {(x-z_2)^2_\perp\over x^+}+i\epsilon\Big]^3}
\langle \Tr\{U_{z_1}U^\dagger_{z_2}\}\rangle_A
\nonumber
\end{eqnarray} 
where we also used $\tr\{\gamma^-\hatY_{2\perp}\gamma^+\hatX_{2\perp}\}= 4(X_2,Y_2)_\perp$.
Note that we indicated with $z_2$ the point at which the shock wave cuts the quark propagator.
The point $z_1$, instead, is the point at which the shock-wave cut the straight dotted blue line which represents the gauge-link, and it is a
point that runs from point $x$ to point $y$. 

The evolution of the trace of two infinite Wilson lines with respect to the rapidity is the BK equation. 
However, for our purpose, it will be enough the linearization of the BK eq. i.e. the BFKL equation which we can write as 
\begin{eqnarray}
&&\hspace{-2cm}2a{d\over da}\,\calv^a(z_\perp) 	
= {\alpha_s N_c\over \pi^2}\!\int\!d^2z'
\Big[{\calv^a(z'_\perp)\over (z-z')_\perp^2} - {(z,z')_\perp\calv^a(z_\perp)\over z'^2_\perp(z-z')^2_\perp}\Big]\,,
\label{calvevolution}
\end{eqnarray}
with
\begin{eqnarray}
&&{1\over z^2_\perp}\calu(z_\perp) \equiv \calv(z_\perp) \,,
\label{def-calv-nu}
\end{eqnarray}
where $\calu(z_\perp)$ is the forward matrix element which depends only on its transverse size and it is obtained from
$\calu(x_\perp,y_\perp) = 1-{1\over N_c}\tr\{U(x_\perp)$ $U^\dagger(y_\perp)\}$.
In eq. (\ref{calvevolution}) we made use of the two dimensional scalar product defined as $(x,y)_\perp = x^1y^1+x^2y^2$.

Evolution equation (\ref{calvevolution}) is with respect to the coordinate space evolution parameter defined as
\begin{eqnarray}
a = -{2x^+y^+\over (x-y)^2a_0}+i\epsilon\,,
\label{coordparam}
\end{eqnarray}
which is reminiscent of the composite Wilson lines operators introduced to restore the 
M\"obius conformal invariance lost at NLO level (for details see Refs. 
\cite{Balitsky:2009xg, Balitsky:2009yp, Balitsky:2010ze, Balitsky:2012bs}).
The peculiarity of the evolution parameter $a$ is that it is in coordinate space and it suits well our purpose.

The solution of evolution equation (\ref{calvevolution}) is
\begin{eqnarray}
\calv^a(z_{12}) = \int\!{d\nu\over 2\pi^2}(z_{12}^2)^{-\half+i\nu} \left(a\over a_0\right)^{\aleph(\gamma)\over 2}
\int d^2\omega(\omega^2_\perp)^{-\half -i\nu}\calv^{a_0}(\omega_\perp)\,,
\label{solution}
\end{eqnarray}
where $\aleph(\gamma)\equiv \bar{\alpha}_s\chi(\gamma)$, with $\bar{\alpha}_s = {\alpha_s N_c\over \pi}$,
$\chi(\gamma)=  2\psi(1) - \psi(\gamma) - \psi(1-\gamma)$, and, as usual, $\gamma=\half+i\nu$.
The parameter $a_0$ is the initial point of evolution  which is $a_0={P^-\over M_N}$ with $M_N$ the mass of the hadronic target. 

It is convenient to define the parameters $u={|y^+|\over \Delta^+}$, $\baru = {x^+\over \Delta^+}$,
and use them to rewrite the solution of the BFKL equation, eq. (\ref{solution}) as
\begin{eqnarray}
\calv^a(z_{12})
=  \int{d\nu\over  2\pi^2}
\,(z_{12}^2)^{-\half+i\nu}\left(-{2L^2u\baru\over \Delta_\perp^2}{{P^-}^2\over M^2_N}+i\epsilon\right)^{{\aleph(\gamma)\over 2}}
\!\!\int d^2\omega (\omega^2_\perp)^{-\half-i\nu}\calv^{a_0}(\omega_\perp)
\label{solution2}
\end{eqnarray}
with $\Delta^+ = x^++|y^+| = L$,
$\calv^a(z_{12}) = {1\over z_{12}^2}\calu^a(z_{12})$, and $\Delta = (x-y)$. 

Let us convolute the solution of the evolution equation of forward matrix element, eq. (\ref{solution2}), with eq. (\ref{quark-startinpoint})
and we arrive at
\begin{figure}[t!]
	\begin{center}
		\includegraphics[width=5.5in]{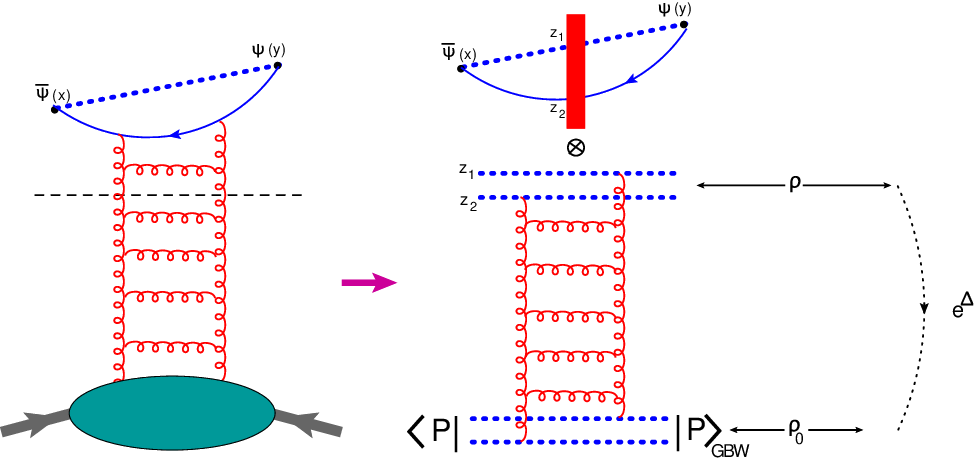}
		\caption{
			In this diagrammatic illustration of the high-energy OPE, the target state is depicted by a green blob in the left panel, while the target is evaluated using the GBW model in the right panel. The dashed black line in the left panel signifies the factorization in rapidity, separating the fields into quantum (upper part of the diagram) and classical (lower part of the diagram). In the right panel, the red band represents the classical field of the target, consisting of gluon fields that undergo Lorentz contraction and time dilation, ultimately forming a shock wave. The resummation of the ladder diagrams, which represent the logarithms of the Ioffe-time parameter, is incorporated through the exponentiation of the Pomeron intercept.
	}
		\label{Fig:quarkLOif-evolution}
	\end{center}
\end{figure}
\begin{eqnarray}
&&\hspace{-1cm}\int_0^{+\infty} dx^+ \int_{-\infty}^0dy^+\delta(x^+-y^+-L)
\langle\barpsi(x^+,x_\perp)\gamma^-[x^+n^\mu+x_\perp, y^+n^\mu+y_\perp]\psi(y^+,y_\perp)\rangle
\nonumber\\
&&\hspace{-1cm}
={i\,N_c\over \pi^3}\!\!\int_0^{+\infty} {dx^+ \over {x^+}^2}\,
\int_{-\infty}^0{dy^+\over {y^+}^2}\delta(x^+-y^+-L)\!\int d^2z_2\,
{(x-z_2)^2_\perp+(y-z_2)^2_\perp - (x-y)^2_\perp\over 
\Big[{(y-z_2)^2_\perp\over |y^+|} + {(x-z_2)^2_\perp\over x^+}+i\epsilon\Big]^3}\langle\calu(z_{12})\rangle_A
\nonumber\\
&&\hspace{-1cm}
= {i\,N_c\over \pi^3}\!\!\int_0^{+\infty} {dx^+ \over {x^+}^2}
\int_{-\infty}^0{dy^+\over {y^+}^2}\delta(x^+-y^+-L)\int d^2z_2
{(x-z_2)^2_\perp+(y-z_2)^2_\perp - (x-y)^2_\perp
\over \Big[{(y-z_2)^2_\perp\over |y^+|} + {(x-z_2)^2_\perp\over x^+}+i\epsilon\Big]^3}z^2_{12}
\nonumber\\
&&\hspace{-0.5cm}
\times\!\!\int{d\nu\over  2\pi^2}
\,(z_{12}^2)^{-\half+i\nu}\left(-{2L^2u\baru\over \Delta_\perp^2}
{{P^-}^2\over M^2_N}+i\epsilon\right)^{{\alpha_s N_c\over 2\pi}\chi(\nu)}
\int d^2\omega (\omega^2_\perp)^{-\half-i\nu}\langle\calv^{a_0}(\omega_\perp)\rangle_{\rm GBW}
\end{eqnarray}
The GBW notation within the angle bracket signifies that the matrix element will be evaluated using 
the GBW model~\cite{Golec-Biernat:1998js}. 
For an alternative model with saturation we can use the McLerran-Venugopalan (MV)~\cite{McLerran:1993ni} model.
In the next section, instead, we will consider a model without saturation mechanism.

We need the projection of the impact factor on the power like eigenfunctions which is
\begin{eqnarray}
&&\hspace{-2cm}\int d^2z {\big[(x-z_2)^2_\perp + (y-z_2)^2_\perp - (x-y)_\perp^2\big] z_{12}^{2\gamma}
\over \Big[{(y-z_2)^2_\perp\over |y^+|} + {(x-z_2)^2_\perp\over x^+}+i\epsilon\Big]^3}
= {2\pi^2 {\Delta^+}^3\over [\Delta^2_\perp]^{1-\gamma}}
{\gamma^2\,(u\baru)^{\gamma+2}\over \sin\pi\gamma}
\label{quark-projection}
\end{eqnarray}
Result (\ref{quark-projection}) is obtained using
${(y-z_2)^2_\perp\over |y^+|} + {(x-z_2)^2_\perp\over x^+}
= {1\over \Delta^+u\baru}\left[(z_2-x_u)^2_\perp + \Delta^2_\perp u\baru\right]$,
where $z_{1\perp}$ is a point which runs from $x_\perp$ to $y_\perp$, so it can be parametrized as
$z_{1\perp} =x_u = ux_\perp + \baru y_\perp$.
In (\ref{quark-projection}) notation $z_{12}^2 = (z_1-z_2)^2_\perp$ is also used.
Using result (\ref{quark-projection}), we have
\begin{eqnarray}
&&\hspace{-1.3cm}\int_0^{+\infty} dx^+ \int_{-\infty}^0dy^+\delta(x^+-y^+-L)
\langle\barpsi(x^+,x_\perp)\gamma^-[x^+n^\mu+x_\perp, y^+n^\mu+y_\perp]\psi(y^+,y_\perp)\rangle
\nonumber\\
&&\hspace{-1.3cm}
= {i\,N_c\over \pi^3}\int_0^{+\infty} {dx^+ \over {x^+}^2}
\int_{-\infty}^0 {dy^+\over {y^+}^2}\,\delta(x^+-y^+-L)
\nonumber\\
&&\hspace{-0.8cm}
\times\!\int \!d\nu\, { {\Delta^+}^3\gamma^2\over [\Delta^2_\perp]^{1-\gamma}}
{(u\baru)^{\gamma+2}\over\sin \pi\gamma}
\,\left(-{2L^2u\baru\over \Delta_\perp^2}{{P^-}^2\over M^2_N}+i\epsilon\right)^{\!\!\aleph(\gamma)\over 2}
\!\!\!\int d^2\omega (\omega^2_\perp)^{-\half-i\nu}\langle\calv^{a_0}(\omega_\perp)\rangle_{\rm GBW}
\nonumber\\
&&\hspace{-1.3cm}
=
{i\,N_c\over \pi^2}\!\int \!d\nu\,{\gamma^2[\Delta^2_\perp]^{\gamma-1}\,\over \sin\pi\gamma}
{\Gamma^2(1+\gamma)\over \Gamma(2+2\gamma)}
\,\left(-{2L^2\over \Delta_\perp^2}{{P^-}^2\over M^2_N}+i\epsilon\right)^{\!\!\aleph(\gamma)\over 2}
\!\!\!\int d^2\omega (\omega^2_\perp)^{-\half-i\nu}\langle\calv^{a_0}_\omega\rangle_{\rm GBW} 
\label{quark-midpoint}
\end{eqnarray}
where we used
\begin{eqnarray}
&&\hspace{-2cm}\int_0^1 du\,(u\baru)^{\gamma+{\aleph(\gamma)\over 2}} 
= {\Gamma^2(1+\gamma+{\aleph(\gamma)\over 2})\over \Gamma(2+2\gamma+\aleph(\gamma))}
\simeq {\Gamma^2(1+\gamma)\over \Gamma(2+2\gamma)} + \calo(\alpha_s)
\label{q-uint}
\end{eqnarray}
The $\alpha_s$ correction omitted in equation (\ref{q-uint}) would contribute if we were to also compute the next-to-leading order (NLO) impact factor and employ the solution of the NLO BFKL equation,\textit{ i.e.}, perform a comprehensive NLO calculation.

As mentioned above, we will use the GBW model to evaluate the forward dipole matrix element
\begin{eqnarray}
	\bra{P}\calu(x)\ket{P+\epsilon^-n'} =&& P^-2\pi\delta(\epsilon^-)\langle\bra{P}\calu(x)\ket{P}\rangle\nonumber\\
&&=P^-2\pi\delta(\epsilon^-) \sigma_0\left(1-\exp\left({-x^2_\perp Q^2_s\over 4}\right)\right)
\label{GBWmodel}
\end{eqnarray}
where $Q_s$ is the saturation scale and $\sigma_0=29.12 \,{\rm mb}$ is the dimension of the dipole cross section whose numerical 
value was obtained from fitting Hera data~\cite{Golec-Biernat:1998js}. 
In high-energy QCD, the saturation scale $Q_s$ is particularly important when studying the behavior of hadronic matter at small Bjorken-x. 
The saturation scale is related to the gluon density inside hadrons and offers a quantitative measure of the transition between the linear regime of parton evolution and the non-linear regime dominated by parton saturation. 
In our case, we can view the saturation scale as a non-perturbative parameter. 

Using model (\ref{GBWmodel}) in eq. (\ref{quark-midpoint}) and integrating over $\omega_\perp$ we arrive at
\begin{eqnarray}
&&\hspace{-2cm}\int_0^{+\infty}\!\! dx^+ \!\!\int_{-\infty}^0dy^+\delta(x^+-y^+-L)
{1\over 2P^-}\langle P|\barpsi(x^+,x_\perp)\gamma^-
\nonumber\\
&&\hspace{-2cm}
\times[x^+n^\mu+x_\perp, y^+n^\mu+y_\perp]\psi(y^+,y_\perp)|P\rangle
\nonumber\\
&&\hspace{-2cm}
= {i\,N_c\sigma_0\over 2\pi^3 \Delta^2_\perp}
\int\! d\nu\,
{\Gamma^3(1+\gamma)\over \Gamma(2+2\gamma)}\Gamma^2(1-\gamma)
\left(-{2L^2\over \Delta_\perp^2}{{P^-}^2\over M^2_N}+i\epsilon\right)^{{\aleph(\gamma)\over 2}}
\left({Q^2_s\Delta^2_\perp\over 4}\right)^\gamma\,.
\label{q-ancont}
\end{eqnarray}
Now, we can finally rewrite result (\ref{q-ancont}) 
using the space like vector $z^\mu$, $z^2<0$, and obtain the high-energy behavior of the operator defined in eq. (\ref{defoperator})
\begin{eqnarray}
\hspace{-1cm}\calm(\varrho,z^2)
= {i\,N_c\sigma_0\over 2\pi|z|^2}\int d\nu
\left({2\varrho ^2\over z^2M^2_N}+i\epsilon\right)^{{\aleph(\gamma)\over 2}}
{\gamma^3\over \sin^2(\pi\gamma)}{\Gamma(\gamma)\over\Gamma(2+2\gamma)}
\left({Q^2_s|z|^2\over 4}\right)^\gamma
\label{quark-bilocalresult-z}
\end{eqnarray}
As previously mentioned, at high energy, we do not distinguish the $0$ component from the $3$ 
component, allowing us to define the Ioffe-time parameter as $\varrho = z \cdot P = LP^-$, since $(P^-)^2 = (P \cdot \frac{z}{|z|})^2
={\varrho^2\over -z^2}$.

The evaluation of the integral over the parameter $\nu$ in result (\ref{quark-bilocalresult-z}) can be performed numerically or
in the saddle point approximation.
In eq. (\ref{quark-bilocalresult-z}),  
the factor ${\gamma^3\over \sin^2(\pi\gamma)}{\Gamma(\gamma)\over\Gamma(2+2\gamma)}$ 
is a slowly varying function of $\nu$. Thus, in the saddle point approximation, eq. (\ref{quark-bilocalresult-z}) is
\begin{eqnarray}
\hspace{-1cm} \calm(\varrho,z^2) \simeq {i\,N_c\over 64}{Q_s\sigma_0\over|z|}
\left({2\varrho^2\over z^2 M^2_N}+i\epsilon\right)^{\bar{\alpha}_s 2\ln 2}
{e^{-{\ln^2 {Q_s|z|\over 2}\over 
7\zeta(3)\bar{\alpha}_s\ln\left({2\varrho^2\over z^2 M^2_N}+i\epsilon\right)}}
\over\sqrt{7\zeta(3)\bar{\alpha}_s\ln\left({2\varrho^2\over z^2 M^2_N}+i\epsilon\right)}}\,.
\label{q-saddpoint}
\end{eqnarray}
The outcome presented in eq. (\ref{q-saddpoint}) embodies the characteristic logarithm resummation, featuring the exponentiation of the well-known Pomeron intercept within the saddle point approximation. In this particular instance, the resummed logarithms correspond to the large values of the Ioffe-time parameter $\rho$.

\begin{figure}[t!]
	\begin{center}
		\includegraphics[width=2.5in]{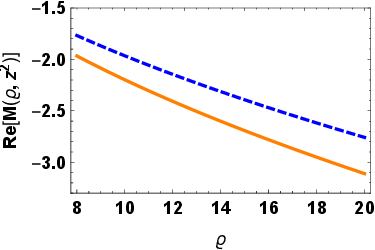}
		\includegraphics[width=2.5in]{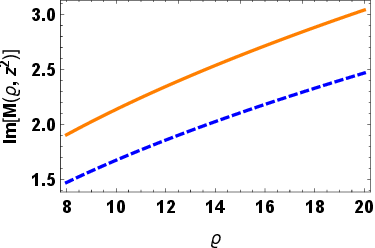}
		\caption{In the left and right panel we plot the real and imaginary part, respectively, of the Ioffe-time amplitude; we compare the numerical
	evaluation of eq. (\ref{quark-bilocalresult-z}) (orange curve) with its saddle point approximation, eq. (\ref{q-saddpoint}) (blue dashed curve).
To obtain the plots we used  $|z|=0.5$, and $M_N=1$ GeV.}
		\label{Fig:IofAmpNumSadlReIm}
	\end{center}
\end{figure}

In Fig. \ref{Fig:IofAmpNumSadlReIm}, we have plotted both the real and imaginary parts of the quark 
Ioffe-time distribution using the saddle point approximation, as given by equation (\ref{q-saddpoint}) 
(represented by the blue dashed curve), and the numerical evaluation of equation (\ref{quark-bilocalresult-z}) (depicted by the orange curve).
The plots were obtained using $Q_s=(x_0/x_B)^{0.277\over 2}$ with $x_0=0.41\times 10^{-4}$~\cite{Golec-Biernat:1998js}.
However, for the Ioffe-time amplitude, we will use $Q_s=(x_0\varrho_0)^{0.277\over 2}$ where $\varrho_0$
is the starting point of the evolution, which, as can be observed from Fig. \ref{Fig:IofAmpNumSadlReIm}, is $\varrho_0=8$,
so $Q_s=0.33 \, {\rm GeV}$. The value of the strong coupling we used is $\bar{\alpha}_s = {\alpha_s N_c\over \pi} = 0.2$.

\section{Ioffe-time distribution at leading twist and next-leading twist}
\label{sec:Ioffetwist}

In the previous section we have obtained the behavior of the Ioffe-time distribution in the BFKL limit. 
As it is known, the BFKL limit captures the behavior of the structure function at all twist. To better understand this statement,
let us rewrite result (\ref{q-ancont}) as an integral along the imaginary axes by changing the integration variable from parameter $\nu$,
to $\gamma=\half+i\nu$ 
and take its Mellin transform, thus obtaining
\begin{eqnarray}
&&\hspace{-0.6cm}\int_{\Delta^2_\perp M_N}^{+\infty}dL\, L^{-j}\int_0^{+\infty} dx^+ \int_{-\infty}^0dy^+\delta(x^+-y^+-L)
\label{quarkMellin}\\
&&\times\langle\barpsi(x^+,x_\perp)\gamma^-[x^+n^\mu+x_\perp, y^+n^\mu+y_\perp]\psi(y^+,y_\perp)\rangle
\nonumber\\
&&\hspace{-0.6cm}
={N_c\over \pi \Delta^2_\perp}\int_{\half-i\infty}^{\half+i\infty} \!\! d\gamma\int_{\Delta^2_\perp M_N}^{+\infty}\!\! dL\, L^{-j+\aleph(\gamma)}
\!\!\left({Q^2_s\Delta^2_\perp\over 4}\right)^\gamma\!\!\!
{\gamma^3\over \sin^2(\pi\gamma)}{\Gamma(\gamma)\over\Gamma(2+2\gamma)}\!\!
\left(\!-{2\over \Delta_\perp^2}{{P^-}^2\over M^2_N}+i\epsilon\right)^{{\!\!\! \aleph(\gamma)\over 2}}
\nonumber
\end{eqnarray}
Performing the Mellin transform we have
\begin{eqnarray}
&&\int_{\Delta^2_\perp M_N}^{+\infty}dL\, L^{-j}\int_0^{+\infty} dx^+ \int_{-\infty}^0dy^+\delta(x^+-y^+-L)
\nonumber\\
&&\times\langle\barpsi(x^+,x_\perp)\gamma^-[x^+n^\mu+x_\perp, y^+n^\mu+y_\perp]\psi(y^+,y_\perp)\rangle
\nonumber\\
&&={N_c\over \pi\Delta^2_\perp}\int_{\half-i\infty}^{\half+i\infty} \!\! d\gamma\,
\theta(\Re[\omega - \aleph(\gamma)]){(\Delta^2_\perp M_N)^{-\omega + \aleph(\gamma)}\over \omega - \aleph(\gamma)}
\,\left({Q^2_s\Delta^2_\perp\over 4}\right)^\gamma
\nonumber\\
&&\times
{\gamma^3\over \sin^2(\pi\gamma)}{\Gamma(\gamma)\over\Gamma(2+2\gamma)}
\,\left(\!-{2\over \Delta_\perp^2}{{P^-}^2\over M^2_N}+i\epsilon\right)^{{\aleph(\gamma)\over 2}}
\label{quarkMellin1}
\end{eqnarray}
The integration over $\gamma$ can be calculated by taking the residue closing the contour to the right
because $0<{Q^2_s\Delta^2_\perp\over 4}<1$.
If we indicate with $\tilde{\gamma}$ the solution of $\omega-\aleph(\tilde{\gamma})=0$, we get 
\begin{eqnarray}
&&\int_{\Delta^2_\perp M_N}^{+\infty}dL\, L^{-j}\int_0^{+\infty} dx^+ \int_{-\infty}^0dy^+\delta(x^+-y^+-L)
\nonumber\\
&&\times\langle\barpsi(x^+,x_\perp)\gamma^-[x^+n^\mu+x_\perp, y^+n^\mu+y_\perp]\psi(y^+,y_\perp)\rangle
\nonumber\\
&&={2 i N_c\over \Delta^2_\perp \aleph'(\tilde{\gamma})}
\left({Q^2_s\Delta^2_\perp\over 4}\right)^{\!\!\tilde{\gamma}}\!
{\tilde{\gamma}^3\over \sin^2(\pi\tilde{\gamma})}{\Gamma(\tilde{\gamma})\over\Gamma(2+2\tilde{\gamma})}
\,\left(\!-{2\over \Delta_\perp^2}{{P^-}^2\over M^2_N}+i\epsilon\right)^{{\omega\over 2}}
\label{quarkMellin2}
\end{eqnarray}
Where, we recall, $\omega=j-1$.
If we take the inverse Mellin transform of (\ref{quarkMellin2}), we would reproduce again the result of the Ioffe-time amplitude
in the BFKL approximation that we already obtained in the previous section. 

If we relax the BFKL condition ${\alpha_s\over \omega} \simeq 1$, 
and consider the limit $\alpha_s\ll\omega\ll1$, we can demonstrate that the result (\ref{quarkMellin2}) 
can be computed by resumming an infinite series of residues, with each contribution having a higher power of $\Delta$. 
Since we consider $0<\Delta^2_\perp<1$, the series consists of contributions that become less significant 
as the power of $\Delta^2_\perp$ increases. 
This is essentially a twist expansion in coordinate space. Our objective is then to calculate the first two twist contributions, 
which are the ones most likely to be computed within the lattice formalism.

To proceed let us observe that
\begin{eqnarray}
\aleph(\gamma) = \bar{\alpha}_s\left(2\psi(1) - \psi(\gamma) - \sum\limits_{n=1}^{N}{1\over n-\gamma}-\psi(N+1-\gamma)\right)\,,
\label{alephexpand}
\end{eqnarray} 
In the limit $\gamma\to 1$ we have 
$\aleph(\gamma) \to {\bar{\alpha}_s\over 1-\gamma}$ and 
${1\over \aleph(\gamma)-\omega}\to {1\over {\bar{\alpha}_s\over 1-\gamma}-\omega} = 
{1-\gamma\over \omega(\gamma -1 + {\bar{\alpha}_s\over \omega})}$. Taking the residue at 
$\gamma=1-{\bar{\alpha}_s\over \omega}$, from eq. (\ref{quarkMellin1}) we have
\begin{eqnarray}
&&
\int_{x^2_\perp M_N}^{+\infty}\!\!dL\, L^{-j}
{1\over 2P^-}\langle P|\bar{\psi}(L,x_\perp)\gamma^-[nL+x_\perp,0]\psi(0) |P\rangle
\nonumber\\
&&= {iN_cQ^2_s \sigma_0\over 24\pi^2\bar{\alpha}_s}
\,\left({Q^2_sx^2_\perp\over 4}\right)^{\!-{\bar{\alpha}_s\over \omega}}
\!\left(\!-{2\over x_\perp^2}{{P^-}^2\over M^2_N}+i\epsilon\right)^{\!\!{\omega\over 2}}
\label{q-analytic-res1}
\end{eqnarray}
where we used the small $\alpha_s$ expansion for
\begin{eqnarray}
{\gamma^3\over \sin^2(\pi\gamma)}{\Gamma(\gamma)\over\Gamma(2+2\gamma)}(1-\gamma)
\stackrel{\gamma=1-{\bar{\alpha}_s\over \omega}}{\simeq} {\omega\over 6\pi^2\bar{\alpha}_s}
+ \calo(\bar{\alpha}_s^0)
\label{approxyGammaratio}
\end{eqnarray}
The next-to-leading residue, \textit{i.e.} the next-to-leading contribution in terms of $Q_s^2 x^2_\perp$ expansion for
$x^2_\perp\to 0$,
is at $\gamma=2-{\bar{\alpha}_s\over \omega}$. Indeed, as done for the leading residue, in the limit of $\bar{\alpha}_s\ll\omega\ll 1$,
it is easy to show that we just need to make the substitution $\aleph(\gamma)\to {\bar{\alpha}_s\over 2-\gamma}$
and ${1\over \aleph(\gamma)-\omega}\to {1\over {\bar{\alpha}_s\over 2-\gamma}-\omega} = 
{2-\gamma\over \omega(\gamma -2 + {\bar{\alpha}_s\over \omega})}$. So from eq. (\ref{quarkMellin1}) we take the residue at
$\gamma=2-{\bar{\alpha}_s\over \omega}$  and obtain
\begin{eqnarray}
&&\hspace{-2cm}
\int_{x^2_\perp M_N}^{+\infty}dL\, L^{-j}\, 
{1\over 2P^-}\langle P|\bar{\psi}(L,x_\perp)\gamma^-[nL+x_\perp,0]\psi(0) |P\rangle
\nonumber\\
&&\hspace{-2cm}
= {iN_cQ^2_s \sigma_0\over 24\pi^2\bar{\alpha}_s}
\,\left({Q^2_sx^2_\perp\over 4}\right)^{-{\bar{\alpha}_s\over \omega}}
\,\left(\!-{2\over x_\perp^2}{{P^-}^2\over M^2_N}+i\epsilon\right)^{{\omega\over 2}}
{Q^2_sx^2_\perp\over 5}\,,
\label{q-analytic-res2}
\end{eqnarray}
where expanding in $\bar{\alpha}_s$ we used
\begin{eqnarray}
{\gamma^3\over \sin^2(\pi\gamma)}{\Gamma(\gamma)\over\Gamma(2+2\gamma)}(2-\gamma)
\stackrel{\gamma=2-{\bar{\alpha}_s\over \omega}}{\simeq} {\omega\over 15\pi^2\bar{\alpha}_s}
+ \calo(\bar{\alpha}_s^0)\,.
\label{approxyGammaratio2}
\end{eqnarray}

Besides the series of the moving (dynamical) poles of which we calculated the first two leading ones, there is also a non moving pole
at $\gamma=1$. The contribution of this pole does not contribute as explain in the Appendix and as also demonstrated 
in the gluon case~\cite{Chirilli:2021euj}.

Summing the two leading residues, results (\ref{q-analytic-res1}) and (\ref{q-analytic-res2}),
we finally arrive at 
\begin{eqnarray}
&&\hspace{-2cm}
\int_{x^2_\perp M_N}^{+\infty}dL\, L^{-j}\, 
{1\over 2P^-}\langle P|\bar{\psi}(L,x_\perp)\gamma^-[nL+x_\perp,0]\psi(0) |P\rangle
\nonumber\\
&&\hspace{-2cm}
= {iN_cQ^2_s \sigma_0\over 24\pi^2\bar{\alpha}_s}
\,\left({Q^2_sx^2_\perp\over 4}\right)^{\!-{\bar{\alpha}_s\over \omega}}
\!\! \left(\!-{2\over x_\perp^2}{{P^-}^2\over M^2_N}+i\epsilon\right)^{\!{\omega\over 2}}
\left(1 + {Q^2_sx^2_\perp\over 5}\right)\,.
\label{q-2leadingres-sum}
\end{eqnarray}
The two leading residues we just calculated organize themselves as an
expansion in $x^2_\perp$: the sub-leading term is suppressed by an extra power of $x^2_\perp$ which is typical of the
coordinate space twist expansion.

We now need to perform the inverse Mellin transform. 
We will consider only the case $0<{Q_s^2x^2_\perp\over 4}<1$ 
which is consistent with twist expansion. Indeed, in performing the inverse Mellin transform, one could also consider the 
case ${Q_s^2x^2_\perp\over 4}>1$, but this case is inconsistent with \textit{twist expansion} because higher powers
of $x^2_\perp$ cannot be disregarded.
The inverse Mellin transform of (\ref{q-2leadingres-sum}) is
\begin{eqnarray}
&&\hspace{-2cm}
{1\over 2\pi i}\int_{1-i\infty}^{1+i\infty} d\omega\, L^{\omega}
{iN_cQ^2_s \sigma_0\over 24\pi^2\bar{\alpha}_s}
\,\left({Q^2_s|z|^2\over 4}\right)^{\!-{\bar{\alpha}_s\over \omega}}
\,\left({2\over z^2}{{P^-}^2\over M^2_N}+i\epsilon\right)^{\!{\omega\over 2}}
\left(1 + {Q^2_s|z|^2\over 5}\right)
\nonumber\\
&&\hspace{-2cm}
=  {iN_cQ^2_s \sigma_0\over 12\pi^2\bar{\alpha}_s}
\left({\bar{\alpha}_s\ln{2\over Q_s|z|}|\over \ln\left({2\varrho^2\over z^2M^2_N}+i\epsilon\right)}\right)^\half 
\left(1 + {Q^2_s|z|^2\over 5}\right)I_1(u) + \calo\left({Q^4_s|z|^4\over 16}\right) 
\label{LT-NLTioffeamp}
\end{eqnarray}
where $I_1(u)$ is the modified Bessel function with
\begin{eqnarray}
u=\left[4\bar{\alpha}_s\ln\left({2\over Q_s|x_\perp|}\right)\ln\left({2\varrho^2\over z^2M^2_N}+i\epsilon\right)\right]^\half
\end{eqnarray}
\begin{figure}[t!]
	\begin{center}
		\includegraphics[width=2.8in]{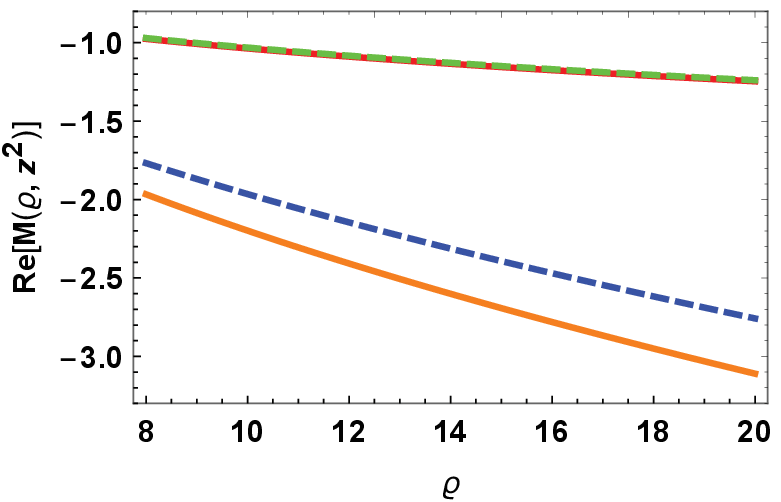}
		\includegraphics[width=2.8in]{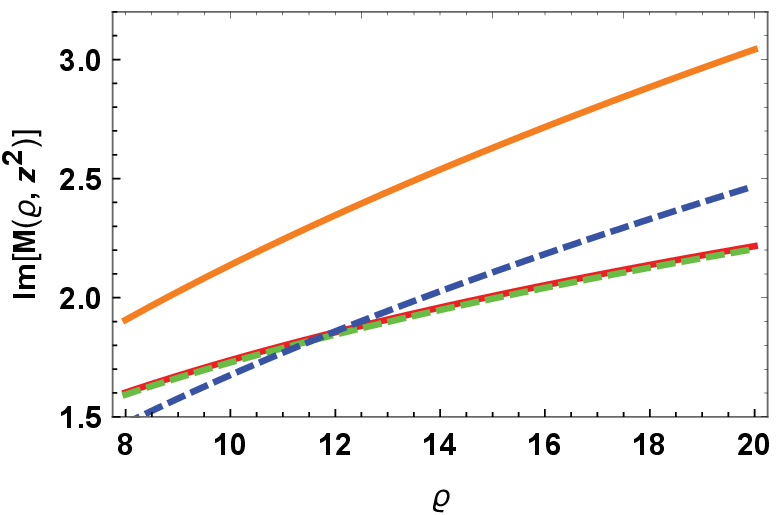}
		\caption{In the left and right panel we plot the real and imaginary part, respectively of the Ioffe-time amplitude; we compare the numerical
			evaluation of eq. (\ref{quark-bilocalresult-z}) (orange curve) with its saddle point approximation, eq. (\ref{q-saddpoint}) with
			the LT, eq. (\ref{q-analytic-res1}) (green dashed curve), and the NLT (\ref{q-2leadingres-sum}) (red solid curve).}
		\label{Fig:IofAm-twistSadBFKLReIm}
	\end{center}
\end{figure}
The result given in eq. (\ref{LT-NLTioffeamp}) represents the Ioffe-time amplitude, including next-to-leading twist corrections. 
In Fig. \ref{Fig:IofAm-twistSadBFKLReIm}, we show the real and imaginary components of the leading (eq. (\ref{q-saddpoint}), green dashed curve) and next-to-leading (eq. (\ref{q-2leadingres-sum}), solid red curve) twist corrections, in comparison with the BFKL resummation result.

For the real part, the leading and next-to-leading twist corrections exhibit a value of approximately $-1$ at Ioffe-time $\varrho=8$ and display a modest decrease for larger values of $\varrho$. Conversely, the BFKL resummed curves possess a value of roughly $-2$ for $\varrho=8$ and demonstrate a swift decline for higher values of $\varrho$. Regarding the imaginary part, although an overlap between the leading and next-to-leading twist corrections and the numerical outcome of the BFKL resummation (blue dashed curve) is observed at $\varrho=12$, the leading and next-to-leading twist corrections exhibit a gentle increase, while the BFKL resummed result rises more rapidly.

It is noteworthy that the behavior observed for large values of $\varrho$ in the leading and next-to-leading twist, as illustrated in Fig. \ref{Fig:IofAm-twistSadBFKLReIm}, is consistent with the lattice calculation results shown in Fig. 1 of Ref. \cite{Joo:2020spy}
(see also \cite{Joo:2019jct}). 
We can observe a similar consistency also with the results presented in Fig. 2 and 3 of Ref. \cite{Bhat:2022zrw}
even though in \cite{Bhat:2022zrw} the pseudo-Ioffe-time distribution has been calculated for a maximum value the Ioffe-time parameter 
$\varrho = 8$, thus making the comparison (especially for the imaginary part) for large Ioffe-time values a bit difficult.

This consistency suggests that lattice calculations may not adequately capture higher twist contributions.
Indeed, the behavior of the pseudo Ioffe-time distribution with BFKL resummation (all twist resummed) 
represented by the orange solid curve and dashed blue curve in Fig. \ref{Fig:IofAm-twistSadBFKLReIm}
suggests a stepper decrease for the real part and a steeper increase for the imaginary part.

\section{Ioffe-time amplitude with the photon impact factor model}
\label{sec:IoffetimePhoton}

In this section, our main goal lies in demonstrating that the twist expansion, as derived in the prior section, 
is not simply an incidental result of the chosen model. To maintain the focus on the salient findings, 
we confine the intricate details of the computation to the Appendix, while here, we concentrate on the presentation of the primary outcomes. 

\begin{figure}[t!]
	\begin{center}
		\includegraphics[width=5.5in]{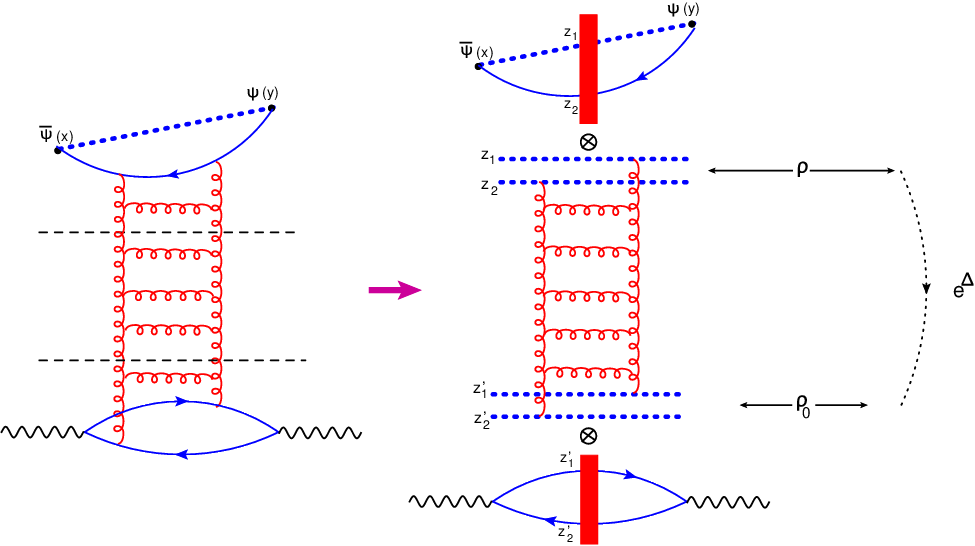}
		\caption{
			Here we diagrammatically illustrate the high-energy OPE with the photon impact factor model.
			The two black dashed lines represent the factorization in rapidity. The diagram at the bottom of the right panel
			is the diagram for the photon impact factor.
		}
		\label{Fig:IoffePhoton}
	\end{center}
\end{figure}

The Ioffe-time amplitude, when considered in relation to the photon impact factor, is as follows (please see Fig. \ref{Fig:IoffePhoton}).
\begin{eqnarray}
&&\hspace{-1cm}
\int dx^+ dy^+\delta(x^+-y^+-L)
\Big\langle\barpsi(x^+,x_\perp)\gamma^-[x^+n^\mu+x_\perp, y^+n^\mu+y_\perp]\psi(y^+,y_\perp)
\nonumber\\
&&\hspace{-1cm}
\times\varepsilon^{\lambda,\,\mu}{\varepsilon^{\lambda,\,\nu}}^*\left(2\over s\right)^{\!\half}\!\!\int\! d^4x' d^4y' \delta(y'^+)
\,e^{iq\cdot (x'-y')}j^\mu(x')j^\nu(y')\Big\rangle
\nonumber\\
&&\hspace{-1cm}
\label{Ioffephotoncorrelation}
\end{eqnarray}
with transverse polarizations $\varepsilon^{\lambda,\mu} = (0,0,\vec{\varepsilon}^\lambda_\perp)$,
$\vec{\varepsilon}^\lambda_\perp = (-1/\sqrt{2})(\lambda,i)$, and $\lambda=\pm1$.
Here $s$ is the large parameter with dimension of energy square (Mandelstam variable).

Applying the high-energy OPE to (\ref{Ioffephotoncorrelation}) as done in the case of the GBW model in the previous section, 
we arrive at (see section \ref{IoffePhotonDetails} for the details of the calculation)
\begin{eqnarray}
&&\hspace{-1cm}
\int dx^+ dy^+\delta(x^+-y^+-L)
\Big\langle\barpsi(x^+,x_\perp)\gamma^-[x^+n^\mu+x_\perp, y^+n^\mu+y_\perp]\psi(y^+,y_\perp)\Big\rangle
\nonumber\\
&&\hspace{-1cm}
\times\Big\langle\varepsilon^{\lambda,\,\mu}{\varepsilon^{\lambda,\,\nu}}^*\left(2\over s\right)^{\!\half}\!\!\int\! d^4x' d^4y' \delta(y'^+)
\,e^{iq\cdot (x'-y')}j^\mu(x')j^\nu(y')\Big\rangle
\nonumber\\
&&\hspace{-1cm}
={i\alpha_s^2N^2_c\over 32 \pi^4|Q||\Delta_\perp|}\!\int\,d\nu\,
{(\gamma\bamma+2)\over (1-\gamma)\gamma}{\Gamma^2(1-\gamma)\over 2\gamma+1}
{\Gamma^3(1+\gamma)\over \Gamma(2+2\gamma)}
{\Gamma^2(\gamma)\over\Gamma(2\gamma)}
\nonumber\\
&&\hspace{-1cm}
\times{\Gamma^3(2-\gamma)\over \Gamma(4-2\gamma)}
\left(Q^2\Delta^2_\perp\over 4\right)^{i\nu}
\left({2L^2\over \Delta_\perp^2}{\sqrt{2}{q^-}^2\over Q^2}\right)^{{\alpha_s N_c\over 2\pi}\chi(\nu)}
\label{photonBFKL}
\end{eqnarray}
where we remind that $\Delta_\perp^2 = (x-y)^2_\perp$, and $-q^2=Q^2>0$.
The initial point of the evolution is $a_0 = {Q^2 \over \sqrt{2}{q^-}^2}$ (see Appendix \ref{IoffePhotonDetails} for details).
We can now evaluate result (\ref{photonBFKL}) in the saddle-point approximation and arrive at
\begin{eqnarray}
&&\hspace{-1cm}
\int dx^+ dy^+\delta(x^+-y^+-L)
\Big\langle\barpsi(x^+,x_\perp)\gamma^-[x^+n^\mu+x_\perp, y^+n^\mu+y_\perp]\psi(y^+,y_\perp)
\nonumber\\
&&\hspace{-1cm}
\times\varepsilon^{\lambda,\,\mu}{\varepsilon^{\lambda,\,\nu}}^*\left(2\over s\right)^{\!\half}\!\!\int\! d^4x' d^4y' \delta(y'^+)
\,e^{iq\cdot (x'-y')}j^\mu(x')j^\nu(y')\Big\rangle
\nonumber\\
&&\hspace{-1cm}
= {i\bar{\alpha}_s^2\over 32 |Q||\Delta_\perp|}{9\pi^3\sqrt{\pi}\over 512}
{e^{-{\ln^2{Q^2\Delta^2_\perp\over 4}}\over
7\zeta(3)\bar{\alpha}_s\ln\left(- {2\sqrt{2}\varrho^2\over \Delta^2_\perp Q^2}+i\epsilon\right) }
\over \sqrt{7\zeta(3)\bar{\alpha}_s\ln\left(- {2\sqrt{2}\varrho^2\over \Delta^2_\perp Q^2}+i\epsilon\right)}}
\left(- {2\sqrt{2}\varrho^2\over \Delta^2_\perp Q^2}+i\epsilon\right)^{\bar{\alpha}_s 2\ln 2}
\label{saddlephoton}
\end{eqnarray}
In the case of photon impact factor the role of the Ioffe-time parameter is
$\varrho=Lq^-$ and the large logarithms resummed are of the type $\ln {\varrho}$ for large $\varrho$.

To obtain the leading and next-to-leading twist correction from (\ref{photonBFKL})
we perform first a Mellin transform, calculate the first two residues and then Mellin transform back and
obtain (see section \ref{IoffePhotonDetails} for the details of the calculation)
\begin{eqnarray}
&&\hspace{-1cm}
{1\over 2\pi i} \int_{1-i\infty}^{1+i\infty} d\omega\, L^\omega
\int_{\Delta^2_\perp \Delta E}^{+\infty}\!dL \, L^{-j}\int dx^+ dy^+\delta(x^+-y^+-L)
\Big\langle\barpsi(x^+,x_\perp)\gamma^-
\nonumber\\
&&\hspace{-1cm}
\times[x^+n^\mu+x_\perp, y^+n^\mu+y_\perp]\psi(y^+,y_\perp)\varepsilon^{\lambda,\,\mu}
{\varepsilon^{\lambda,\,\nu}}^*\left(2\over s\right)^{\!\half}\int d^4x' d^4y' \delta(y'^+)
\,e^{iq\cdot (x'-y')}j_\mu(x')j_\nu(y')\Big\rangle
\nonumber\\
&&\hspace{-1cm}
= {i\,\over 288 \pi} 
{ 2\bar{\alpha}_s \ln{4\over Q^2\Delta^2_\perp}\over 
\ln\left(- {2\sqrt{2}\varrho^2\over \Delta^2_\perp Q^2}+i\epsilon\right)}
I_2(h)\left(1- {3\over 50 }{Q^2\Delta^2_\perp\over 4}\right)
\label{ioffetwistphoton}
\end{eqnarray}
where $I_2(h)$ is the modified Bessel function with
\begin{eqnarray}
h = \left[2\bar{\alpha}_s\ln{4\over Q^2\Delta^2_\perp}
\ln\left(-{2\sqrt{2}\varrho^2\over \Delta^2_\perp Q^2}+i\epsilon\right)\right]^\half\,.
\label{hh}
\end{eqnarray}
\begin{figure}[t!]
	\begin{center}
		\includegraphics[width=2.8in]{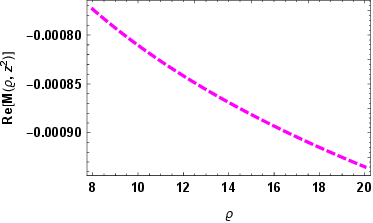}
		\includegraphics[width=2.8in]{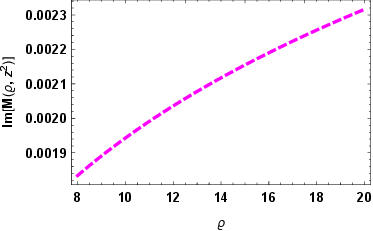}
		\caption{In the left and right panel we plot the real and imaginary part, respectively of the Ioffe-time amplitude
			at the leading twist eq. (\ref{ioffetwistphoton}); the plot is obtained using the photon-impact factor model with
			$Q=0.33$ GeV which is the same value as $Q_s$ at Ioffe-time $\varrho=8$.}
		\label{Fig:Ioffephotontwist}
	\end{center}
\end{figure}
Result (\ref{ioffetwistphoton}) shows that the twist corrections organize themselves as an expansion 
in $Q^2\Delta^2_\perp$ like the one we obtained using the GBW model, thus proving that
this feature is not peculiar to the model GBW model only.
Moreover, from Fig. \ref{Fig:Ioffephotontwist}, where
we plot only the  leading residue (the next-to-leading residue is just a very small shift), we observe that
the behavior of the Ioffe-time amplitude 
at leading twist (and next-to-leading) with the photon impact factor model
is not only similar to the one obtained with the GBW model which is presented in Fig. \ref{Fig:IofAm-twistSadBFKLReIm} but also
quite close to the one obtained through Lattice calculation in ref.~\cite{Joo:2020spy}.

\section{Quark pseudo-PDF}
\label{sec:quarkpseudo}

In this section, we will perform the Fourier transform of the Ioffe-time distribution in the BFKL limit to derive the quark pseudo PDF. 
Following this, we will calculate the same transformation for both leading and next-to-leading twist corrections. Finally, we will compare the pseudo-PDF in the BFKL approximation with the leading and next-to-leading twists by plotting their corresponding values within the $x_B$ range of $[0.01, 0.1]$.
\subsection{Pseudo PDF in the BFKL approximation}

We will perform the Fourier transform of eq. (\ref{quark-bilocalresult-z}) following the pseudo PDF definition. 
With the assumption that $0 \le x_B \le 1$, we have
\begin{eqnarray}
&&\hspace{-1.3cm}\int_{-\infty}^{+\infty}{d\varrho\over 2\pi}\,
e^{-i\varrho x_B}\calm(\varrho,z^2)
\nonumber\\
&&\hspace{-1.3cm}
={i\,N_c\sigma_0\over 2\pi|z|^2}\!\int_{-\infty}^{+\infty}\!{d\varrho\over 2\pi}\,e^{-i\varrho x_B}\!\!\int d\nu\!
\left({2\varrho ^2\over z^2M^2_N}+i\epsilon\right)^{{\aleph(\gamma)\over 2}}\!\!
{\gamma^3\over \sin^2(\pi\gamma)}{\Gamma(\gamma)\over \Gamma(2+2\gamma)}
\left({Q^2_s|z|^2\over 4}\right)^\gamma
\nonumber\\
&&\hspace{-1.3cm}
= - {i\,N_c Q_s\sigma_0 \over 4\pi^2|z|x_B}\int d\nu
\left({2\over z^2 M^2_Nx^2_B}+i\epsilon\right)^{{\aleph(\gamma)\over 2}}
\sin\big[{\pi\over 2}\aleph(\gamma)\big]
\nonumber\\
&&\hspace{-1.3cm}
~~~\times\!{\gamma^2\,\Gamma(1+\aleph(\gamma))\over \sin^2(\pi\gamma)}{\Gamma(1+\gamma)\over \Gamma(2+2\gamma)}
\left({Q^2_s|z|^2\over 4}\right)^{i\nu}\,,
\label{pseudoQpdf}
\end{eqnarray}
with, we recall, $z^2<0$.
Result (\ref{pseudoQpdf}) is the quark pseudo PDF. We can simplify this result further using the approximation 
$\Gamma(1+\aleph(\gamma))\sin[\pi\aleph(\gamma)/2]\simeq {\pi\over 2}\aleph(\gamma)+ \calo(\bar{\alpha}_s)$.
However, in plotting this result we will not take into account such approximation.
 
The integration over the parameter $\nu$ can be evaluated
in saddle point approximation. Using again $\aleph(\gamma=1/2) = \bar{\alpha}_s 4\ln 2$ and
\begin{eqnarray}
{\gamma^2\over \sin^2(\pi\gamma)}{\Gamma(1+\gamma)\over \Gamma(2+2\gamma)}
\stackrel{\gamma=1/2}{=}{\sqrt{\pi}\over 16}\,,
\end{eqnarray}
we have
\begin{figure}[t!]
	\begin{center}
		\includegraphics[width=2.8in]{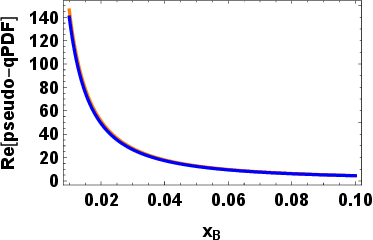}
		\includegraphics[width=2.9in]{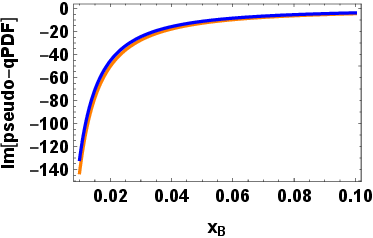}
		\caption{Here we plot the real and imaginary parts of the numerical evaluation of eq. (\ref{pseudoQpdf}) (orange curve) 
			with its saddle point approximation, eq. (\ref{pseudo-quarkBFKL}) (blue curve). }
		\label{Fig:Quarkpseudo-num-vs-sadd}
	\end{center}
\end{figure}
\begin{eqnarray}
\int_{-\infty}^{+\infty}{d\varrho\over 2\pi}\,
e^{-i\varrho x_B}\calm(\varrho,z^2) = \!\!\! && - {i\,N_c Q_s\sigma_0 \over 64\pi|z| |x_B|}
{e^{-{\ln^2{Q_s |z|\over 2}\over 7\zeta(3)\bar{\alpha}_s\ln\left({2\over x^2_Bz^2 M^2_N}+i\epsilon\right)}}
	\over \sqrt{7\zeta(3)\bar{\alpha}_s\ln\left({2\over x^2_Bz^2 M^2_N}+i\epsilon\right)}}
\nonumber\\
\times \!\!\! &&
{\left({2\over x^2_Bz^2 M^2_N}+i\epsilon\right)^{\bar{\alpha}_s2\ln 2}
\over\Big(\Gamma(1+\bar{\alpha}_s4\ln 2)\sin\!\big[2\pi\bar{\alpha}_s\ln 2\big]\Big)^{-1}}\,,
\label{pseudo-quarkBFKL}
\end{eqnarray}
where we can use again $\Gamma[1+\bar{\alpha}_s 4\ln 2]\sin\!\big[2\pi\bar{\alpha}_s\ln 2\big] \simeq 2\pi\bar{\alpha}_s\ln 2 
+ \calo(\bar{\alpha}_s)$. We observe that the behavior of the pseudo-PDF in the BFKL limit is primarily dictated by the well-known exponentiation of the Pomeron intercept, which effectively resums the logarithms of $1/x_B$. 
This point is described in Fig. \ref{Fig:Quarkpseudo-num-vs-sadd}, 
in which we present the results from eq. (\ref{pseudo-quarkBFKL}) along with the numerical evaluation of (\ref{pseudoQpdf}). The comparison demonstrates the efficacy of our saddle point approximation as described in eq. (\ref{pseudo-quarkBFKL}).

\subsection{Pseudo PDF in the leading and next-to-leading approximation}

Let us proceed by calculating the pseudo-PDF in the leading and next-to-leading approximation, that is we perform the Fourier transform of the
leading and next-to-leading twist correction calculated for the Ioffe-time distribution.

It is convenient to perform the inverse Mellin transform after the Fourier transform.
So, our starting point is eq. (\ref{q-2leadingres-sum}) (here $L >0$)
\begin{eqnarray}
&&\hspace{-0.3cm}
\int_{1-i\infty}^{1+i\infty}\!\! {d\omega \over 2\pi i}\, L^{\omega}\!\!
\int_0^{+\infty}{dLP^-\over 2\pi}\,e^{-iLP^-x_B}\int_{x^2_\perp M_N}^{+\infty}dL\, L^{-j}\, 
{1\over 2P^-}\langle P|\bar{\psi}(L,x_\perp)\gamma^-[nL+x_\perp,0]\psi(0) |P\rangle
\nonumber\\
&&\hspace{-0.3cm}
= {iQ^2_s \sigma_0\over 24\pi\alpha_s}\!\int_{1-i\infty}^{1+i\infty} {d\omega\over 2\pi i}\int_0^{+\infty}{dLP^-\over 2\pi}\,e^{-iLP^-x_B}
\left({Q^2_sx^2_\perp\over 4}\right)^{\!-{\bar{\alpha}_s\over \omega}}
\left(\!-{2\over x_\perp^2}{{L^2P^-}^2\over M^2_N}+i\epsilon\!\right)^{\!{\omega\over 2}}
\left(\!1 + {Q^2_sx^2_\perp\over 5}\right)
\nonumber\\
&&\hspace{-0.3cm}
= {Q^2_s \sigma_0\over 48\pi^2\alpha_sx_B}\int_{1-i\infty}^{1+i\infty} {d\omega\over 2\pi i}
\left({Q^2_sx^2_\perp\over 4}\right)^{-{\bar{\alpha}_s\over \omega}}\!\Gamma(1+\omega)
\left(\!{2\over x_\perp^2 x_B^2M^2_N}\right)^{{\omega\over 2}}
\left(1 + {Q^2_sx^2_\perp\over 5}\right)
\label{quarkPseudo-twist}
\end{eqnarray}
Since we are in the approximation $\omega\ll 1$ we can use $\Gamma(1+\omega)\sim 1$ in (\ref{quarkPseudo-twist})
and we obtain
\begin{figure}[t!]
	\begin{center}
		\includegraphics[width=3in]{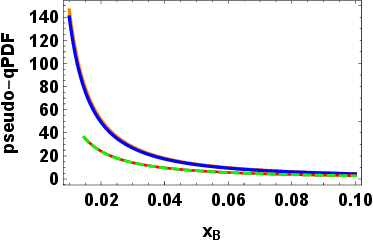}
		\caption{The plot presents the quark pseudo PDF by comparing the numerical evaluation of eq. (\ref{pseudoQpdf}) (illustrated by the orange curve) with its saddle point approximation derived from eq. (\ref{pseudo-quarkBFKL}) (portrayed by the blue curve). Furthermore, we display the LT (marked by the green dashed curve) and the NLT (signified by the solid red curve) obtained from eq. (\ref{quarkPseudo-twist}).}
		\label{Fig:Quarkpseudo-num-vs-sadd-vs-twist}
	\end{center}
\end{figure}
\begin{eqnarray}
&&\hspace{-0.3cm}
\int_{1-i\infty}^{1+i\infty}\!\! {d\omega \over 2\pi i}\, L^{\omega}\!\!
\int_0^{+\infty}{dLP^-\over 2\pi}\,e^{-iLP^-x_B}\int_{x^2_\perp M_N}^{+\infty}dL\, L^{-j}\, 
{1\over 2P^-}\langle P|\bar{\psi}(L,x_\perp)\gamma^-[nL+x_\perp,0]\psi(0) |P\rangle
\nonumber\\
&&\hspace{-0.3cm}
= {Q^2_s \sigma_0\over 24\pi^2\alpha_sx_B}
\left({\bar{\alpha}_s\ln{2\over Q_s|x_\perp|}\over
\ln\left(2\over x^2_\perp x_B^2 M^2_N\right)}\right)^{\!\half}
\!\!\left(1 + {Q^2_sx^2_\perp\over 5}\right)\!I_1(v)\,,
\label{pseudo-quark-twists}
\end{eqnarray}
where $I_1$ is the modified Bessel function and we defined
\begin{eqnarray}
v\equiv \left[ 4\bar{\alpha}_s \ln{2\over Q_s|x_\perp|}\ln\left(2\over x^2_\perp x_B^2 M^2_N\right)\right]^\half\,.
\end{eqnarray}
We again performed the inverse Mellin transform only in the region $0<{Q_s x^2_\perp\over 4}<1$
where the higher twist effects can be considered small corrections.
in Fig. \ref{Fig:Quarkpseudo-twist-vs-twistapproxy}, we compare result (\ref{quarkPseudo-twist})
with eq. (\ref{pseudo-quark-twists}), where the approximation $\Gamma(1+\omega) \simeq 1$ has been employed. 
Our analysis demonstrates the validity of the approximation $\Gamma(1+\omega) \simeq 1$.

\begin{figure}[t!]
	\begin{center}
		\includegraphics[width=3in]{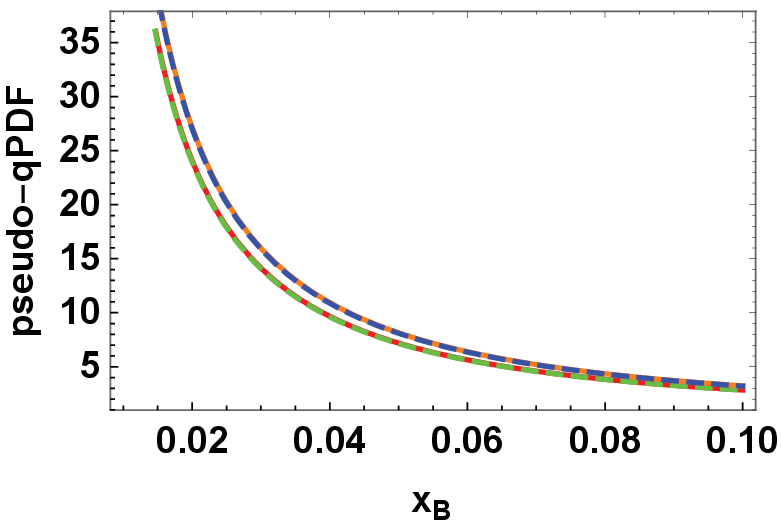}
		\caption{The plot illustrates the quark pseudo PDF for both LT and NLT, considering the cases with and without the approximation $\Gamma(1+\omega) \simeq 1$ for $\omega \ll 1$. The blue and orange curves represent the results obtained without employing the approximation $\Gamma(1+\omega) \simeq 1$.}
		\label{Fig:Quarkpseudo-twist-vs-twistapproxy}
	\end{center}
\end{figure}

In Fig. \ref{Fig:Quarkpseudo-num-vs-sadd-vs-twist}, we observe that the first two twist contributions,
which have only the real part, as described by eq. (\ref{pseudo-quark-twists}), 
exhibit strong agreement with the all-twists resummed BFKL results (\ref{pseudo-quarkBFKL}) and 
 (\ref{pseudoQpdf}) for larger values of $x_B$.
However, as we approach smaller $x_B$ values, the two twist corrections deviate from the full BFKL result, suggesting that,
lattice calculations, now available only at large values of $x_B$, may describe well only the region where leading twist contributions dominate, 
which is the domain of DGLAP dynamics. On the other hand, the small-$x_B$ region, where all twist corrections contribute equally and 
are described by the BFKL dynamics, is not attainable by lattice calculations.

\section{Quark quasi-PDF}
\label{sec:quarkquasi}

\subsection{Quark quasi-PDF in the BFKL approximation}

In this section we are going to obtain the quark quasi-PDF by performing the Fourier transform of the Ioffe-time distribution
given in eq. (\ref{quark-bilocalresult-z}). To this end, following Ref.~\cite{Chirilli:2021euj}, 
we introduce the real parameter $\varsigma$, with $-z^2=\varsigma^2>0$, and
the four-vector $\xi^\mu\equiv {z^\mu\over |z|} = {z^\mu\over |\varsigma|}$ with $|z|=\sqrt{-z^2}$.
Under the large boost we can then identify $(LP^-)^2 = (z_\mu P^\mu)^2 = \varsigma^2P_\xi^2$. 
So, we can perform the following substitution
\begin{eqnarray}
\left({\varrho^2\over z^2 M^2_N}+i\epsilon\right)^{\aleph(\gamma)\over 2} 
\to \left(-{P^2_\xi\over M^2_N}+i\epsilon\right)^{\aleph(\gamma)\over 2}\,.
\end{eqnarray}

Therefore, starting from (\ref{defoperator}) and (\ref{quark-bilocalresult-z}), and using ${\ssz\over 2z\cdot p} = {1\over 2P_\xi}\,\ssxi $,
the quasi-distribution, \textit{i.e.} the pseudo Ioffe-time distribution rewritten \'a la quasi-PDF is
\begin{eqnarray}
{1\over 2P_\xi}\bra{P}\barpsi(\varsigma){\ssxi}[\varsigma, 0]\psi(0)\ket{P}
=&& {i\,N_c Q_s\sigma_0 \over 4\pi |\varsigma|}\int d\nu
\left(-{2P_\xi^2\over M^2_N}+i\epsilon\right)^{{\aleph(\gamma)\over 2}}
\nonumber\\
\times&&{\gamma^3\over \sin^2(\pi\gamma)}{\Gamma(\gamma)\over \Gamma(2+2\gamma)}
\left({Q^2_s|\varsigma|^2\over 4}\right)^{i\nu} + \calo(\alpha_s)\,.
\label{quasi-distribu}
\end{eqnarray}
\begin{figure}[t!]
	\begin{center}
		\includegraphics[width=2.8in]{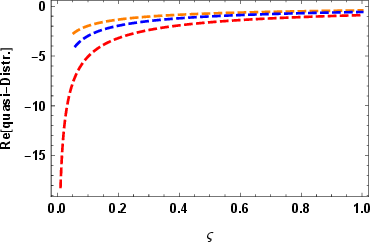}
		\includegraphics[width=2.7in]{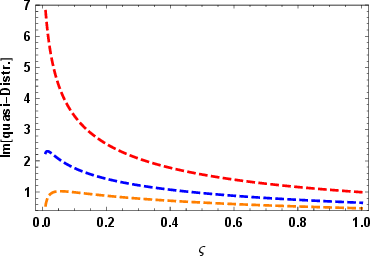}
		\caption{The plot illustrates the quark quasi-Distribution, eq. (\ref{quasi-distribu}) for two different values of
		$P_\xi$. The red dashed curve is obtained with $P_\xi=10$ GeV, the blue dashed one with $P_\xi=4$ GeV, 
		and the orange dashed one with $P_\xi=2$ GeV. For the numerical evaluation we also used $\bar{\alpha}_s=0.2$,
		$Q_s=0.33$ GeV, and $M_N =1$ GeV.}
		\label{Fig:quasidistribu}
	\end{center}
\end{figure}
In Fig. \ref{Fig:quasidistribu} we plot the quasi-Distribution, eq. (\ref{quasi-distribu}), 
for $P_\xi=10$ GeV (red dashed curve), $P_\xi=4$ GeV (Blue dashed curve), and $P_\xi=2$ GeV (orange dashed curve).
Result presented in Fig. \ref{Fig:quasidistribu} can be compared with the ones
obtained, although with smaller values of $P_\xi$, in lattice calculations in refs.~\cite{Alexandrou:2015rja, Alexandrou:2016jqi}.

Evaluating the integration over the $\nu$ parameter in the saddle point approximation the quasi-distribution (\ref{quasi-distribu})
becomes
\begin{eqnarray}
\hspace{-1cm}
{1\over 2P_\xi}\bra{P}\barpsi(\varsigma){\ssxi}[\varsigma, 0]\psi(0)\ket{P}
\simeq {iN_c Q_s\sigma_0\over 64 |\varsigma|}
\left(-{2P_\xi^2\over M^2_N}+i\epsilon\right)^{\!\bar{\alpha}_s2\ln 2}\!\!
{e^{-{\ln^2{Q_s|\varsigma|\over 2}\over 7\zeta(3)\bar{\alpha}_s\ln\left(-{2P_\xi^2\over M^2_N}+i\epsilon\right)}}
\over \sqrt{7\zeta(3)\bar{\alpha}_s\ln\left(-{2P_\xi^2\over M^2_N}+i\epsilon\right)}}\,.
\label{quasi-distribuSaddle}
\end{eqnarray}

\begin{figure}[t!]
	\begin{center}
		\includegraphics[width=2.9in]{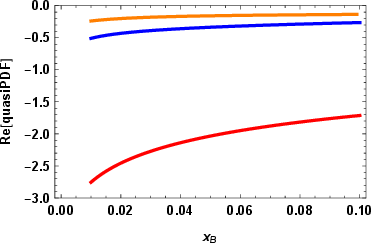}
		\includegraphics[width=2.9in]{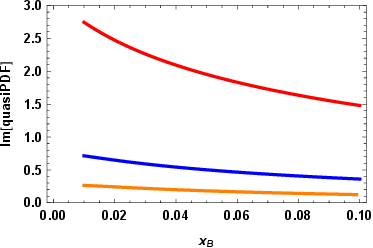}
		\caption{In the left and the right panel we plot, respectively, the real and the imaginary part of the numerical evaluation of eq. (\ref{quark-quasiPDFnum}) with $P_\xi=10$ GeV (red curve), $P_\xi=4$ GeV (blue curve), $P_\xi=2$ GeV (orange curve). 
			For the numerical evaluation we used $\bar{\alpha}_s=0.2$, $M_N=1$ GeV.
			and $Q_s=0.33$ GeV.}
		\label{Fig:quark-quasiPDFreim}
	\end{center}
\end{figure}

The quark quasi-PDF is obtained performing the Fourier transform with respect to the $\varsigma$ parameter.
We notice that, contrary to the pseudo-PDF case, if we perform the Fourier transform of eq. (\ref{quasi-distribu}) first, and then
the integration with respect to the $\nu$ parameter, we end up with a divergent integral.
So, in this case we will proceed as follow. We first integrate the $\nu$ parameter in the saddle point approximation, and then perform the
Fourier transform. So, the numerical evaluation of the Fourier transform of eq. (\ref{quasi-distribuSaddle}) is
\begin{eqnarray}
&&\hspace{-0.3cm}P_\xi\int_{-\infty}^{+\infty} {d\varsigma\over 2\pi} \, e^{-i\varsigma P_\xi x_B}
{1\over 2P_\xi}\bra{P}\barpsi(\varsigma){\ssxi}[\varsigma, 0]\psi(0)\ket{P}
\nonumber\\
&&\hspace{-0.3cm} = \!\!\int_{-\infty}^{+\infty} \!\! d\varsigma \, e^{-i\varsigma P_\xi x_B}
{i\,N_c P_\xi\over 8\pi^2 |\varsigma|}Q_s\sigma_0\!\!\int d\nu
\left(-{2P_\xi^2\over M^2_N}+i\epsilon\right)^{\!\!{\aleph(\gamma)\over 2}}\!
{\gamma^3\Gamma(\gamma)\over \sin^2(\pi\gamma)\Gamma(2+2\gamma)}
\left({Q^2_s|\varsigma|^2\over 4}\right)^{i\nu} + \calo(\alpha_s)
\nonumber\\
&&\hspace{-0.3cm} \simeq {i\,N_c P_\xi Q_s\sigma_0\over 128\pi }
\left(-{2P_\xi^2\over M^2_N}+i\epsilon\right)^{\!\bar{\alpha}_s2\ln 2}
\!\!\int_{-\infty}^{+\infty}\!\! {d\varsigma\over |\varsigma|} \, e^{-i\varsigma P_\xi x_B}
{e^{-{\ln^2{Q_s|\varsigma|\over 2}\over 7\zeta(3)\bar{\alpha}_s\ln\left(-{2P_\xi^2\over M^2_N}+i\epsilon\right)}}
\over \sqrt{7\zeta(3)\bar{\alpha}_s\ln\left(-{2P_\xi^2\over M^2_N}+i\epsilon\right)}}
\label{quark-quasiPDFnum}
\end{eqnarray} 
In Fig. \ref{Fig:quark-quasiPDFreim}, we exhibit the numerical evaluation of eq. (\ref{quark-quasiPDFnum})
with $P_\xi=10$ GeV (red curve), $P_\xi=4$ GeV (blue curve) and $P_\xi=2$ GeV\footnote{The allowed values of $P_\xi$ are the ones such that 
$\bar{\alpha}_s\ln \left({2P_\xi^2\over M_N^2}\right) \sim 1$. The value of $P_\xi=2$ GeV, although barely 
satisfies this condition, it is used because it is a value attainable in lattice calculations.
One should also consider that plots in Fig. \ref{Fig:quark-quasiPDFreim} are obtained using saddle point approximation which
works better at large values of $P_\xi$.} (orange curve), 
demonstrating the distinctive behavior of the quark quasi-PDF. It is crucial to emphasize that this behavior significantly deviates from the quark pseudo-PDF, as represented in Fig. \ref{Fig:Quarkpseudo-num-vs-sadd-vs-twist}. This divergence can be ascribed to the lack of the typical 
logarithm resummation that is a hallmark of the BFKL formalism. We should also enphasize that the saddle point approximation
is a valid approximation for large values of $P_\xi$.

\subsection{Quasi PDF in the leading and next-to-leading approximation}

To obtain the quasi-PDF in the leading and next-to-leading approximation, we follow a similar approach as employed for the pseudo-PDF.
Thus, our starting point is eq. (\ref{q-2leadingres-sum}), which we rewrite using the notation specific to the quasi-PDF
as we did in the previous subsection.
The result (\ref{q-2leadingres-sum}) must undergo an inverse Mellin transformation and a Fourier transformation in the quasi-PDF definition, i.e., maintaining the orientation of the $z^\mu$ vector fixed. 

First, let us obtain the quasi-distribution in the LT and NLT as inverse Mellin transform of eq. (\ref{q-2leadingres-sum})
which is (recall that $\omega=j-1$):
\begin{eqnarray}
&&\hspace{-1cm}{1\over 2\pi i}\int_{1-i\infty}^{1+i\infty}d\omega\,
\varsigma^\omega\int_{x^2_\perp M_N}^{+\infty} d\varsigma\,{\varsigma'}^{-j}
{1\over 2P_\xi}\bra{P}\bar{\psi}(\varsigma')\ssxi[\varsigma',0]\psi(0)\ket{P}
\nonumber\\
&&\hspace{-1cm} = {i N_c Q^2_s \sigma_0\over 24\pi^2 \bar{\alpha}_s}{1\over 2\pi i}\int_{1-i\infty}^{1+i\infty}\!\! d\omega
\left({Q^2_s\varsigma^2\over 4}\right)^{\!-{\bar{\alpha}_s\over \omega}}
\,\left(-{2P_\xi^2\over M^2_N}+i\epsilon\right)^{\!{\omega\over 2}}\!\!\!
\left(1 + {Q^2_s\varsigma^2\over 5}\right)
\nonumber\\
&&\hspace{-1cm}
={i N_c Q^2_s \sigma_0\over 24\pi^2 \bar{\alpha}_s}
\left({4\bar{\alpha}_s\ln{2\over Q_s|\varsigma|}\over \ln\left(-{2P_\xi^2\over M_N^2}+i\epsilon\right)}\right)^\half
I_1(\tildeu)\,,
\label{quasiDistrTwist}
\end{eqnarray}
with
\begin{eqnarray}
\tildeu = \left[4\bar{\alpha}_s\ln{2\over Q_s|\varsigma|}\ln\left(-{2P_\xi^2\over M_N^2}+i\epsilon\right)\right]^\half\,.
\end{eqnarray}
The Inverse Mellin transform (\ref{quasiDistrTwist}) has been performed only in the region $0<Q^2_s\varsigma^2<1$ which is consistent with 
higher twist expansion. 
\begin{figure}[t!]
	\begin{center}
		\includegraphics[width=2.9in]{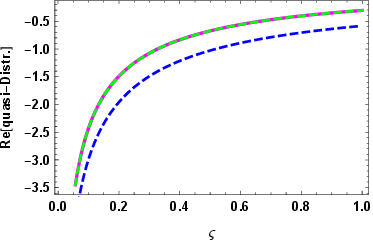}
		\includegraphics[width=2.8in]{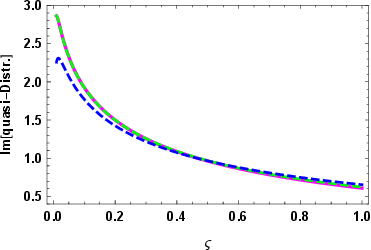}
		\caption{Here we present the plots for the LT (magenta curve) and NLT (green dashed curve), eq. (\ref{quasiDistrTwist}),
			and compare them with the BFKL result (blue dashed curve), eq. (\ref{quasi-distribu}).
			In the left panel we plot the real parts, while in the right panel we plot the imaginary parts.
			The plots are obtained using $P_\xi=4$ GeV, $\bar{\alpha}_s=0.2$, $M_N=1$ GeV.
			and $Q_s=0.33$ GeV.}
		\label{Fig:quasidistribuTwist}
	\end{center}
\end{figure}

In Fig. \ref{Fig:quasidistribuTwist} we plot the real and imaginary parts of the LT and NLT of eq. (\ref{quasiDistrTwist}), and
compare them with the BFKL result eq. (\ref{quasi-distribu}). We notice that the first two leading twist contributions
(the green and magenta curves which are one on top of the other), are consistent with the BFKL result (resummation of
large logarithms of $P_\xi$) within the region $\varsigma \in [0.01,1]$.

The quark quasi-PDF in the LT and NLT is obtained performing the Fourier transform of (\ref{quasiDistrTwist})
with respect to the $\varsigma$ parameter. So we have
\begin{eqnarray}
&&\hspace{-1cm}P_\xi\int_0^{+\infty} {d\varsigma'\over 2\pi}
\, e^{-i\varsigma' P_\xi x_B}{1\over 2\pi i}\int_{1-i\infty}^{1+i\infty}d\omega\,
{\varsigma'}^\omega\int_{x^2_\perp M_N}^{+\infty} d\varsigma\,\varsigma^{-j}
{1\over 2P_\xi}\bra{P}\bar{\psi}(\varsigma)\xi_\mu\gamma^\mu[\varsigma,0]\psi(0)\ket{P}
\nonumber\\
&&\hspace{-1cm} = {iQ^2_sP_\xi \sigma_0\over 48\pi^2\alpha_s}{1\over 2\pi i}\int_{1-i\infty}^{1+i\infty}\!\! d\omega
\int_0^{+\infty} \!\! d\varsigma \, e^{-i\varsigma P_\xi x_B}\left({Q^2_s\varsigma^2\over 4}\right)^{\!-{\bar{\alpha}_s\over \omega}}
\,\left(-{2P_\xi^2\over M^2_N}+i\epsilon\right)^{\!{\omega\over 2}}\!\!\!
\left(1 + {Q^2_s\varsigma^2\over 5}\right)
\nonumber\\
&&\hspace{-1cm} = {iQ^2_s \sigma_0\over 48\pi^2\alpha_s}{1\over 2\pi i}\int_{1-i\infty}^{1+i\infty}\!\! d\omega
P_\xi\left({Q^2_s\over 4}\right)^{-{\bar{\alpha}_s\over \omega}}
\,\left(-{2P_\xi^2\over M^2_N}+i\epsilon\right)^{{\omega\over 2}}
\nonumber\\
&&\hspace{-1cm}~~~\times\!\!
\left({\Gamma(1-{2\alpha_s\over \omega})\over (iP_\xi x_B)^{1-{2\alpha_s\over \omega}}} + 
{\Gamma(3-{2\alpha_s\over \omega})\over (iP_\xi x_B)^{3-{2\alpha_s\over \omega}}}{Q^2_s\over 5}\right)\,.
\end{eqnarray}
We should recall that now we are working in the $\alpha_s\ll\omega\ll 1$ approximation, which means that
we are slightly stepping away from the BFKL limit to enter in the DGLAP one.
To perform the Inverse Mellin we have to distinguish two ranges of the values $x_B$.
For $x_B<{Q_s\over M_N\sqrt{2}}$ we have
\begin{eqnarray}
&&P_\xi\int_0^{+\infty} {d\varsigma'\over 2\pi}
\, e^{-i\varsigma' P_\xi x_B}{1\over 2\pi i}\!\int_{1-i\infty}^{1+i\infty}d\omega\,{\varsigma'}^\omega
\!\!\int_{x^2_\perp M_N}^{+\infty} d\varsigma\,\varsigma^{-j} {1\over 2P_\xi}
\bra{P}\bar{\psi}(\varsigma)\xi_\mu\gamma^\mu[\varsigma,0]\psi(0)\ket{P}
\nonumber\\
\nonumber\\
&& = {N_cQ^2_s \sigma_0\over 48\pi^3\bar{\alpha}_s}{1\over x_B}{1\over 2\pi i}\int_{1-i\infty}^{1+i\infty}\!\! d\omega
\left(-{4P^2_\xi x_B^2\over Q^2_s} + i\epsilon\right)^{\bar{\alpha}_s\over \omega}
\,\left(-{2P_\xi^2\over M^2_N}+i\epsilon\right)^{{\omega\over 2}}
\left(1 - {2Q^2_s\over 5P^2_\xi x_B^2}\right) + \calo\left(\alpha_s\over \omega\right)
\nonumber\\
&& = -{N_cQ^2_s \sigma_0\over 48\pi^3\bar{\alpha}_s}{1\over x_B}
\left({2\bar{\alpha}_s\ln\left(-{Q^2_s\over 4 P^2_\xi x_B^2}-i\epsilon\right)\over
\ln\left(-{2P^2_\xi\over M^2_N}+i\epsilon\right)}\right)^\half\!\! J_1(t)
\left(1 - {2Q^2_s\over 5P^2_\xi x_B^2}\right)+ \calo\left(\alpha_s\over \omega\right)
\label{quarkQuasi-twist-a}
\end{eqnarray}
with
\begin{eqnarray}
t = \left[2\bar{\alpha}_s\ln\left(-{Q^2_s\over 4 P^2_\xi x_B^2}-i\epsilon\right)\ln\left(-{2P^2_\xi\over M^2_N}+i\epsilon\right)\right]^\half
\end{eqnarray}
While for $x_B>{Q_s\over M_N\sqrt{2}}$ we have
\begin{eqnarray}
&&P_\xi\int_0^{+\infty} {d\varsigma'\over 2\pi}
\, e^{-i\varsigma' P_\xi x_B}{1\over 2\pi i}\!\int_{1-i\infty}^{1+i\infty}d\omega\,{\varsigma'}^\omega
\!\!\int_{x^2_\perp M_N}^{+\infty} d\varsigma\,\varsigma^{-j} {1\over 2P_\xi}
\bra{P}\bar{\psi}(\varsigma)\xi_\mu\gamma^\mu[\varsigma,0]\psi(0)\ket{P}
\nonumber\\
\nonumber\\
&& = {N_cQ^2_s \sigma_0\over 48\pi^3\bar{\alpha}_s}{1\over x_B}{1\over 2\pi i}\int_{1-i\infty}^{1+i\infty}\!\! d\omega
\left(-{4P^2_\xi x_B^2\over Q^2_s} + i\epsilon\right)^{\bar{\alpha}_s\over \omega}
\,\left(-{2P_\xi^2\over M^2_N}+i\epsilon\right)^{{\omega\over 2}}
\left(1 - {2Q^2_s\over 5P^2_\xi x_B^2}\right) + \calo\left(\alpha_s\over \omega\right)
\nonumber\\
&& = {N_cQ^2_s \sigma_0\over 48\pi^3\bar{\alpha}_s}{1\over x_B}
\left({2\bar{\alpha}_s\ln\left(-{Q^2_s\over 4 P^2_\xi x_B^2}-i\epsilon\right)\over
\ln\left(-{2P^2_\xi\over M^2_N}+i\epsilon\right)}\right)^\half\!\! I_1(t)
\left(1 - {2Q^2_s\over 5P^2_\xi x_B^2}\right)+ \calo\left(\alpha_s\over \omega\right)
\label{quarkQuasi-twist-b}
\end{eqnarray}
\begin{figure}[t!]
	\begin{center}
		\includegraphics[width=2.9in]{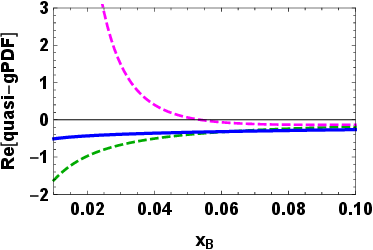}
		\includegraphics[width=2.9in]{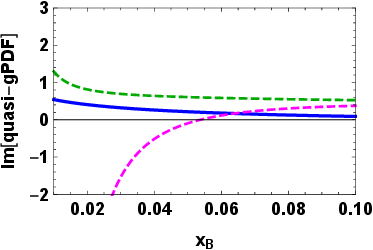}
		\caption{In the left panel we plot the real part of the LT (dark green dashed curve) and NLT (magenta dashed curve)
			of the quasi-PDF distribution eq. (\ref{quarkQuasi-twist-a}) and BFKL resummation (blue curve) 
			eq. (\ref{quark-quasiPDFnum}). In the right panel we plot the imaginary parts.
		The plots are obtained using $P_\xi=4$ GeV, $\bar{\alpha}_s=0.2$, $M_N=1$ GeV, and $Q_s=0.33$ GeV.}
		\label{Fig:Quarkquasi-twist}
	\end{center}
\end{figure}
In Fig. \ref{Fig:Quarkquasi-twist}, we present the leading and next-to-leading twist contributions, as described by 
eqs. (\ref{quarkQuasi-twist-a}) and (\ref{quarkQuasi-twist-b}), for the quasi quark PDF as well as the 
BFKL resummed result eq. (\ref{quarkQuasi-twist-a}). The integral in 
eqs. (\ref{quarkQuasi-twist-a}) and (\ref{quarkQuasi-twist-b})
is performed by closing the contour to the left. To this end, we have 
to impose $\ln{2P_\xi^2\over M^2_N} - \pi >0$. The plots for the LT and NLT corrections
in Fig. \ref{Fig:Quarkquasi-twist} are obtained using 
$P_\xi=4$ GeV. A different choice, such as $P_\xi=2$ GeV, would lead to a divergent integral. 
This means that the formalism we adopted to study the small-$x_B$ behavior of the twist corrections imposes a 
limit on the allowed values of $P_\xi$.
The behavior of the quasi-PDF at LT and NLT is particularly noteworthy due to two primary factors. 
Firstly, it exhibits a stark deviation from the pseudo-PDF case, and secondly, the next-to-leading twist correction displays a divergent trend that opposes the direction of the leading one at small-$x_B$. 
This unusual behavior is attributable to the peculiar nature of the higher twist corrections, 
which are characterized by the $1/(x_B^2 P^2_\xi)$ factor. Indeed, such higher twist corrections, instead of being suppressed as was hoped,
are enhanced at small-$x_B$. We also notice that the full BFKL result is consistent with the LT correction.

\section{Conclusions}
\label{sec:conclusions}

By employing the high-energy operator product expansion, we have derived the high-energy behavior of the Ioffe-time distribution for the quark non-local operator. This finding bears considerable importance for lattice calculations, which are not optimally suited for computing large Ioffe-time behavior. In order to obtain the pseudo distribution from lattice calculations, an extrapolation for extensive Ioffe-time values is required prior to conducting the Fourier transform.

In Fig. \ref{Fig:IofAm-twistSadBFKLReIm}, we presented the Ioffe-time behavior, demonstrating that the Ioffe-time amplitude 
at leading and next-to-leading twist corrections exhibits a large Ioffe-time parameter behavior that aligns with the results 
obtained through lattice calculations in Ref. \cite{Joo:2020spy}. Nonetheless, our analysis also indicates that the leading 
and next-to-leading twist corrections are insufficient to accurately characterize the behavior determined from the complete BFKL result.
Indeed, the higher twist corrections become more important at larger Ioffe-time parameters. While
lattice calculations are, in principle, capable of capturing all twist effects, their restriction to lower values of Ioffe-time 
parameters prohibit them from adequately capturing the domain where the higher twist effects bear the same weight as the leading ones. 
Consequently, lattice calculations limited to low Ioffe-time parameters likely
remain valid only for the first two twist effects which align well with the dynamics as described by DGLAP.
To strengthen our findings, we evaluated a model without saturation effects in Section \ref{sec:IoffetimePhoton}
(see Fig. \ref{Fig:Ioffephotontwist}). The assessment confirmed that both the pattern of the twist expansion and the behavior of the first two leading twists at high Ioffe-time parameters are not exclusive to the GBW model.

Subsequently, we performed an explicit Fourier transform for both the pseudo-PDF and quasi-PDF from the Ioffe-time amplitude. The pseudo-PDF, demonstrated in Fig. \ref{Fig:Quarkpseudo-num-vs-sadd-vs-twist}, follows the expected result in the saddle point approximation, capturing the BFKL resummation (resummation of all twists) and exhibiting an expected rising behavior for small $x_B$ values. Conversely, the quasi-PDF resulted in a different and unusual behavior, reaffirming that they are unsuited for small-$x_B$ studies as was previously observed in the gluon case~\cite{Chirilli:2021euj}.

We also examined the behavior of the pseudo-PDF and quasi-PDF within the first two leading twist contributions, 
which are potentially attainable from lattice calculations. Although the pseudo-PDF's next-to-leading twist contribution exhibits 
similar behavior to that with BFKL resummation, it does not approach the BFKL resummation result significantly faster than the leading contribution.
This observation confirms that an appropriate small-$x_B$ behavior cannot be achieved solely through DGLAP dynamics; 
rather, an all-twist (BFKL) resummation is required.
Indeed, although lattice calculations may in principle capture all twist contributions, the inability to study them in the small-$x_B$ region, where the
higher twist are as important as the leading one, make lattice calculations relevant only for the leading twist effect which are
described by DGLAP dynamics.
Furthermore, while the equivalence between the pseudo-PDF and quasi-PDF formalisms can be demonstrated at moderate $x_B$ values, 
the situation differs for small-$x_B$. Our findings show that the higher twist contributions in the quasi-PDF case are not suppressed 
but instead enhanced, scaling as ${\cal O}(1/(x_B^2 P_\xi^2))$, as shown in equations (\ref{quarkQuasi-twist-a}) and (\ref{quarkQuasi-twist-b}). 
Consequently, this equivalence between the two formalisms is not expendable to include small $x_B$ values.
We also studied the quasi-Distribution (see Fig. \ref{Fig:quasidistribuTwist}), and 
we established that at LT and NLT it is in good agreement with the full BFKL resummed result.
A similar conclusion can be drawn for the 
quasi-PDF displayed in Fig. \ref{Fig:Quarkquasi-twist}. Indeed, although the NLT correction diverges from the LT one for reasons previously discussed,
the behavior of the LT, instead, is in good agreement with the full BFKL resummation.

\section{Acknowledgments}

The author is grateful to Ian Balitsky and Vladimir Braun for valuable discussions.
The author has received financial support from Xunta de Galicia (Centro singular de investigaci\'on de Galicia
accreditation 2019-2022, ref. ED421G-2019/05), by European Union ERDF, by the "Mar\'{\i}a de Maeztu" Units of Excellence program MDM2016-0692, and by the Spanish Research State Agency under project PID2020-119632GB-I00. This work has been performed in the framework of the European Research Council project ERC-2018-ADG-835105 YoctoLHC and the MSCA RISE 823947 "Heavy
ion collisions: collectivity and precision in saturation physics" (HIEIC), and has received funding from the European
Union's Horizon 2020 research and innovation program under grant agreement No. 824093.
I would also like to acknowledge financial support by the High 
Energy Theory Group at the University of Regensburg where this work was started.

\appendix

\section{Derivation of the Ioffe-time amplitude with the photon impact factor model}
\label{IoffePhotonDetails}

In this section we are going to provide more details on the derivation of the Ioffe-time amplitude with the photon
impact factor (IF) model we presented in section \ref{sec:IoffetimePhoton}. 
The Ioffe-time amplitude with photon IF model can be written as

\begin{eqnarray}
&&\hspace{-1cm}
\int dx^+ dy^+\delta(x^+-y^+-L)
\Big\langle\barpsi(x^+,x_\perp)\gamma^-[x^+n^\mu+x_\perp, y^+n^\mu+y_\perp]\psi(y^+,y_\perp)
\nonumber\\
&&\hspace{-1cm}
\times\varepsilon^{\lambda,\,\mu}{\varepsilon^{\lambda,\,\nu}}^*\left(2\over s\right)^{\!\half}\!\!\int\! d^4x' d^4y' \delta(y'^+)
\,e^{iq\cdot (x'-y')}j^\mu(x')j^\nu(y')\Big\rangle\,.
\label{Ioffephotoncorrelation2}
\end{eqnarray}
We will consider virtual photon with transverse polarizations $\varepsilon^{\lambda,\mu} = (0,0,\vec{\varepsilon}^\lambda_\perp)$,
$\vec{\varepsilon}^\lambda_\perp = (-1/\sqrt{2})(\lambda,i)$, and $\lambda=\pm1$.
Here $s$ is the large parameter with dimension of energy square.

The high-energy limit in coordinate space is obtained considering the limits $x^+,y^+\to \infty$ and 
$x'^-,y'^-\to \infty$. Consequently, we can rewrite (\ref{Ioffephotoncorrelation2}) in a factorized form~\cite{Balitsky:2009yp}
\begin{eqnarray}
&&\hspace{-1cm}
\int dx^+ dy^+\delta(x^+-y^+-L)
\Big\langle\barpsi(x^+,x_\perp)\gamma^-[x^+n^\mu+x_\perp, y^+n^\mu+y_\perp]\psi(y^+,y_\perp)\Big\rangle
\nonumber\\
&&\hspace{-1cm}
\times\Big\langle\varepsilon^{\lambda,\,\mu}{\varepsilon^{\lambda,\,\nu}}^*\left(2\over s\right)^{\!\half}\!\!\int\! d^4x' d^4y' \delta(y'^+)
\,e^{iq\cdot (x'-y')}j^\mu(x')j^\nu(y')\Big\rangle
\nonumber\\
&&\hspace{-1cm}
\label{Ioffephotoncorrelationfact}
\end{eqnarray}
We may now apply the high-energy OPE to the two factors in (\ref{Ioffephotoncorrelationfact}). 
We will use the intermediate result (\ref{quark-midpoint}) for the top part of the diagram in Fig. \ref{Fig:IoffePhoton},
and for the photon impact factor, bottom part of Fig. \ref{Fig:IoffePhoton} we use result of section 
\ref{sec:IF} and arrive at
\begin{eqnarray}
&&\hspace{-1cm}
\int dx^+ dy^+\delta(x^+-y^+-L)
\Big\langle\barpsi(x^+,x_\perp)\gamma^-[x^+n^\mu+x_\perp, y^+n^\mu+y_\perp]\psi(y^+,y_\perp)\Big\rangle
\nonumber\\
&&\hspace{-1cm}
\Big\langle\varepsilon^{\lambda,\,\mu}{\varepsilon^{\lambda,\,\nu}}^*\left(2\over s\right)^{\!\half}
\!\!\int\! d^4x' d^4y' \delta(y'^+)
\,e^{iq\cdot (x'-y')}j^\mu(x')j^\nu(y')\Big\rangle
\nonumber\\
&&\hspace{-1cm}
= {i\,N_c\over 2\pi^2}\!\int{d\nu'\over  2\pi^2}{\gamma'\,\Gamma(1-\gamma')\Gamma(1+\gamma')\over [\Delta^2_\perp]^{1-\gamma'}}
{4\Gamma^2(1+\gamma')\over \Gamma(2+2\gamma')}
\nonumber\\
&&\hspace{-1cm}
\times\left({2L^2\over \Delta_\perp^2}{\sqrt{2}q^-\over Q^2}\right)^{{\alpha_s N_c\over 2\pi}\chi(\nu')}
\int d^2\omega (\omega^2_\perp)^{-\half-i\nu'}\calv^{a_0}(\omega_\perp)
\nonumber\\
&&\hspace{-1cm}
\times N_c\!\int {d^2\nu\over \pi^3}
{\Gamma(\bamma)\Gamma^3(2-\gamma)\over \Gamma(4-2\gamma)\Gamma(2+\gamma)}
{2\gamma-1\over 2\gamma+1}{\Gamma^2(\bamma)\over\Gamma(2\bamma)}
{(Q^2)^{\gamma-1}\Gamma(2+\gamma)\over 4^{\gamma+1}}
(\gamma\bamma+2)\tilde{\calu}(\nu)\,,
\label{ioffephoto}
\end{eqnarray}
where in this case $\gamma' = \half + i\nu'$, and $\tilde{\calu}(\nu)$ is defined in eq. (\ref{calu-nu}).
Notice that in the photon IF case, the initial point of the evolution, $a_0$ of the evolution parameter (\ref{def-calv-nu}), 
of the BFKL equation (\ref{calvevolution}) has changed. Indeed we have
\begin{eqnarray}
a^{GBW}_0 = {M_N^2\over {P^-}^2}, ~~~~~~ a^{IF}_0 = {\Delta E\over q^-} = {Q^2\over \sqrt{2}{q^-}^2}\,,
\end{eqnarray}
where $\Delta E = {Q^2\over \sqrt{2}q^+}$ is the inverse of the life-time of the quark-anti-quark 
pair which fluctuates from the virtual photon.
The projection of the dipole operator $\calv(\omega_\perp)$ onto to the LO eigenfunctions is
\begin{eqnarray}
\int d^2\omega (\omega^2_\perp)^{-\half-i\nu}\calv^{a_0}(\omega_\perp)
&&= (\omega^2_\perp)^{-\half-i\nu}{1\over \omega^2_\perp}\calu^{a_0}(\omega_\perp)
\nonumber\\
&&= 2\pi{\Gamma(1-2\gamma)\Gamma^2(\gamma)\over\Gamma^2(1-\gamma)\Gamma(2\gamma)}\, \nu^2\,\calu(-\nu)\,.
\label{Vforward-evolution}
\end{eqnarray}
So, eq. (\ref{ioffephoto}) becomes
\begin{eqnarray}
&&\hspace{-1cm}
\int dx^+ dy^+\delta(x^+-y^+-L)
\Big\langle\barpsi(x^+,x_\perp)\gamma^-[x^+n^\mu+x_\perp, y^+n^\mu+y_\perp]\psi(y^+,y_\perp)\Big\rangle
\nonumber\\
&&\hspace{-1cm}
\Big\langle\varepsilon^{\lambda,\,\mu}{\varepsilon^{\lambda,\,\nu}}^*\left(2\over s\right)^{\!\half}\!\!\int\! d^4x' d^4y' \delta(y'^+)
\,e^{iq\cdot (x'-y')}j^\mu(x')j^\nu(y')\Big\rangle
\nonumber\\
&&\hspace{-1cm}
= {i\,N_c^2\over \pi}\!\int{d\nu'\over  2\pi^2}{\gamma'\,\Gamma(1-\gamma')\Gamma(1+\gamma')\over [\Delta^2_\perp]^{1-\gamma'}}
{4\Gamma^2(1+\gamma')\over \Gamma(2+2\gamma')}
{\Gamma(1-2\gamma')\Gamma^2(\gamma')\over\Gamma^2(1-\gamma')\Gamma(2\gamma')}\, \nu'^2
\nonumber\\
&&\hspace{-1cm}
\times \int {d^2\nu\over \pi^3}
{\Gamma(\bamma)\Gamma^3(2-\gamma)\over \Gamma(4-2\gamma)\Gamma(2+\gamma)}
{2\gamma-1\over 2\gamma+1}{\Gamma^2(\bamma)\over\Gamma(2\bamma)}
{(Q^2)^{\gamma-1}\Gamma(2+\gamma)\over 4^{\gamma+1}}
(\gamma\bamma+2)
\nonumber\\
&&\hspace{-1cm}
\times\left({2L^2\over \Delta_\perp^2}{\sqrt{2}q^-\over Q^2}\right)^{\bar{\alpha}_s\aleph(\gamma')}
\langle\calu(-\nu')\tilde{\calu}(\nu)\rangle
\label{ioffephoto2}
\end{eqnarray}
We need the dipole-dipole amplitude at LO which in the Mellin space is
\begin{eqnarray}
\langle\calu(-\nu')\tilde{\calu}(\nu)\rangle = && - {4\pi^2(N^2_c-1)\over N^2_c}{\alpha_s^2\over \nu^2(1+4\nu^2)^2}\delta(\nu-\nu')
\label{LOcalu-nu-squared}
\end{eqnarray}
So, using (\ref{Vforward-evolution}) and the LO dipole-dipole scattering (\ref{LOcalu-nu-squared})we arrive at
\begin{eqnarray}
&&\hspace{-1cm}
\int dx^+ dy^+\delta(x^+-y^+-L)
\Big\langle\barpsi(x^+,x_\perp)\gamma^-[x^+n^\mu+x_\perp, y^+n^\mu+y_\perp]\psi(y^+,y_\perp)\Big\rangle
\nonumber\\
&&\hspace{-1cm}
\times\Big\langle\varepsilon^{\lambda,\,\mu}{\varepsilon^{\lambda,\,\nu}}^*\left(2\over s\right)^{\!\half}\!\!\int\! d^4x' d^4y' \delta(y'^+)
\,e^{iq\cdot (x'-y')}j^\mu(x')j^\nu(y')\Big\rangle
\nonumber\\
&&\hspace{-1cm}
= {i\alpha_s^2(N^2_c-1)\over 32 \pi^4|Q||\Delta_\perp|}\!\int\,d\nu
{(\gamma\bamma+2)\over (1-\gamma)\gamma}{\Gamma^2(1-\gamma)\over 2\gamma+1}
{\Gamma^3(1+\gamma)\over \Gamma(2+2\gamma)}
{\Gamma^2(\gamma)\over\Gamma(2\gamma)}
\nonumber\\
&&\hspace{-1cm}
\times{\Gamma^3(2-\gamma)\over \Gamma(4-2\gamma)}
\left(Q^2\Delta^2_\perp\over 4\right)^{i\nu}
\left({2L^2\over \Delta_\perp^2}{\sqrt{2}{q^-}^2\over Q^2}\right)^{\bar{\alpha}_s\aleph(\gamma)}
\label{IoffePhoton-nu}
\end{eqnarray}

At this point we may proceed similarly to the case GBW model. We take the Mellin transform, calculate the first two leading
residue and then Mellin transform back. 
Starting with the Mellin transform of (\ref{IoffePhoton-nu}) we have

\begin{eqnarray}
&&\hspace{-0.5cm}
\int_{\Delta^2_\perp \Delta E}^{+\infty}\!dL \, L^{-j}\int dx^+ dy^+\delta(x^+-y^+-L)
\langle\barpsi(x^+,x_\perp)\gamma^-[x^+n^\mu+x_\perp, y^+n^\mu+y_\perp]\psi(y^+,y_\perp)
\nonumber\\
&&\hspace{-0.5cm}
\times\varepsilon^{\lambda,\,\mu}{\varepsilon^{\lambda,\,\nu}}^*\left(2\over s\right)^\half\int d^4x' d^4y' \delta(y'^+)
\,e^{iq\cdot (x'-y')}j_\mu(x')j_\nu(y')\rangle
	\nonumber\\
	&&\hspace{-0.5cm}
	= {i\alpha_s^2N^2_c\over 32 \pi^4|Q||\Delta_\perp|}\!\int d\nu\,
	{\theta[\Re(\omega-\aleph(\gamma))]\over \omega-\aleph(\gamma)}
	{(\gamma\bamma+2)\over (1-\gamma)\gamma}{\Gamma^2(1-\gamma)\over 2\gamma+1}
	{\Gamma^3(1+\gamma)\over \Gamma(2+2\gamma)}
	{\Gamma^2(\gamma)\over\Gamma(2\gamma)}
	{\Gamma^3(2-\gamma)\over \Gamma(4-2\gamma)}
	\nonumber\\
	&&\hspace{-0.5cm}
	\times \left(Q^2\Delta^2_\perp\over 4\right)^{i\nu}
	\left({2\over \Delta_\perp^2}{\sqrt{2}{q^-}^2\over Q^2}\right)^{{\alpha_s N_c\over 2\pi}\chi(\nu)}
	(\Delta^2_\perp \Delta E)^{\aleph(\gamma)-\omega}
	\label{usingpho-mell}
\end{eqnarray}

In the limit $\alpha_s\ll\omega\ll1$ use $\aleph(\gamma)\to {\bar{\alpha}_s\over 1-\gamma}$ and
$\gamma\to 1-{\bar{\alpha}_s\over \omega}$, the leading residue is
\begin{eqnarray}
&&\hspace{-1cm}
\int_{\Delta^2_\perp \Delta E}^{+\infty}\!dL \, L^{-j}\int dx^+ dy^+\delta(x^+-y^+-L)
\langle\barpsi(x^+,x_\perp)\gamma^-[x^+n^\mu+x_\perp, y^+n^\mu+y_\perp]\psi(y^+,y_\perp)
\nonumber\\
&&\hspace{-1cm}
\times\varepsilon^{\lambda,\,\mu}{\varepsilon^{\lambda,\,\nu}}^*\int d^4x' d^4y' \delta(y'_*)
\,e^{iq\cdot (x'-y')}j_\mu(x')j_\nu(y')\rangle
\nonumber\\
&&\hspace{-1cm}
\nonumber\\
&&\hspace{-1cm}
= {\alpha_s^2N^2_c\over 32 \pi^4|Q||\Delta_\perp|}\!\int_{\half-i\infty}^{\half+i\infty} d\gamma\,
\theta[\Re(\omega-\aleph(\gamma))]{(\gamma-1)\over \omega(\gamma - 1 + {\bar{\alpha}_s\over \omega})}
{(\gamma\bamma+2)\over (1-\gamma)\gamma}{\Gamma^2(1-\gamma)\over 2\gamma+1}
{\Gamma^3(1+\gamma)\over \Gamma(2+2\gamma)}
\nonumber\\
&&\hspace{-1cm}
\times {\Gamma^2(\gamma)\over\Gamma(2\gamma)}{\Gamma^3(2-\gamma)\over 
\Gamma(4-2\gamma)}\left(Q^2\Delta^2_\perp\over 4\right)^{i\nu}
\left({2\over \Delta_\perp^2}{\sqrt{2}{q^-}^2\over Q^2}\right)^{\aleph(\gamma)\over 2}
(\Delta^2_\perp \Delta E)^{\aleph(\gamma)-\omega}
\nonumber\\
&&\hspace{-1cm}
\simeq - {i\omega\over 288 \pi}
\left(Q^2\Delta^2_\perp\over 4\right)^{-{\bar{\alpha}_s\over \omega}}
\left({2\over \Delta_\perp^2}{\sqrt{2}{q^-}^2\over Q^2}\right)^{\omega\over 2}
\end{eqnarray}
with $\omega=j-1$ and where for $\gamma\to 1-{\bar{\alpha}_s\over \omega}$ we used 
\begin{eqnarray}
{(\gamma\bamma+2)\over \gamma}{\Gamma^2(1-\gamma)\over 2\gamma+1}
{\Gamma^3(1+\gamma)\over \Gamma(2+2\gamma)}
{\Gamma^2(\gamma)\over\Gamma(2\gamma)}
{\Gamma^3(2-\gamma)\over \Gamma(4-2\gamma)}={\omega^2\over 9\bar{\alpha}_s^2} + \calo(\alpha_s)
\end{eqnarray}
The next-to-leading residue is obtain in the same way, we use $\aleph(\gamma)\to {\bar{\alpha}_s\over 2-\gamma}$
and calculate the residue closing the contour to the right. Summing the first two residue 
and performing the inverse Mellin transform  for the relevant case $\ln{Q^2\Delta^2_\perp\over 4}<0$, we get
\begin{eqnarray}
&&\hspace{-0.5cm}
{1\over 2\pi i} \int_{1-i\infty}^{1+i\infty} d\omega\, L^\omega
\int_{\Delta^2_\perp \Delta E}^{+\infty}\!dL \, L^{-j}\int dx^+ dy^+\delta(x^+-y^+-L)
\langle\barpsi(x^+,x_\perp)\gamma^-\nonumber\\
&&\hspace{-0.5cm}
\times[x^+n^\mu+x_\perp, y^+n^\mu+y_\perp]\psi(y^+,y_\perp)
\varepsilon^{\lambda,\,\mu}{\varepsilon^{\lambda,\,\nu}}^*\left(2\over s\right)^\half\int d^4x' d^4y' \delta(y'^+)
\,e^{iq\cdot (x'-y')}j_\mu(x')j_\nu(y')\rangle
\nonumber\\
&&\hspace{-0.5cm}
= - {i\over 288 \pi}\left(1- {3\over 50 }{Q^2\Delta^2_\perp\over 4}\right)
{1\over 2\pi i} \int_{1-i\infty}^{1+i\infty} d\omega \,\omega
\left(4\over Q^2\Delta^2_\perp\right)^{\bar{\alpha}_s\over \omega}
\left({2L^2\over \Delta_\perp^2}{\sqrt{2}{q^-}^2\over Q^2}\right)^{\omega\over 2}
\nonumber\\
&&\hspace{-0.5cm}
= {i\,\over 288 \pi} 
{ 2\bar{\alpha}_s \ln{4\over Q^2\Delta^2_\perp}\over 
	\ln\left(- {2\sqrt{2}\varrho^2\over \Delta^2_\perp Q^2}+i\epsilon\right)}
I_2(h)\left(1- {3\over 50 }{Q^2\Delta^2_\perp\over 4}\right)
\label{ioffetwistphoton2}
\end{eqnarray}
with $h$ defined in (\ref{hh}) and $\varrho = Lq^-$.

\section{Photon Impact factor}
\label{sec:IF}
Using result of ref.~\cite{Balitsky:2012bs}, the T product of two electromagnetic currents can be expanded
in terms of the impact factor and matrix elements of the trace of two Wilson lines. 

\begin{eqnarray}
&&\hspace{-2cm}\left(2\over s\right)^{\!\half}
\!\!\int d^4x d^4y \delta(y^+)\,e^{iq\cdot (x-y)}{\rm T}\{j^\mu(x)j^\nu(y)\}
\nonumber\\
&&\hspace{-2cm}
= N_c\!\int {d^2\nu\over \pi^3}
{\Gamma(\bamma)\Gamma^2(2-\gamma)\Gamma(2-\gamma)\over \Gamma(4-2\gamma)\Gamma(2+\gamma)}
{2\gamma-1\over 2\gamma+1}{\Gamma^2(\bamma)\over\Gamma(2\bamma)}
{(Q^2)^{\gamma-1}\Gamma(2+\gamma)\over 4^{\gamma+1}}
\nonumber\\
&&\hspace{-2cm}
\times\Bigg\{
(\gamma\bamma+2)P^{\mu\nu}_1 + (3\gamma\bamma+2)P^{\mu\nu}_2
\Bigg\}\int d^2z_0\tilde{\calu}_{a_m}(z_0,\nu)
\label{IF}
\end{eqnarray}
where we defined $-q^2=Q^2>0$. Further, we define
\begin{eqnarray}
&&\tilde{\calu}(\nu) \equiv \int dz_0\,\tilde{\calu}(z_0,\nu)
=  {\Gamma(1-2\gamma)\over \Gamma^2(\bamma)}
{\Gamma^2(\gamma)\over \Gamma(2\gamma)}\,
{1\over \pi}
\!\int d^2z (z^2)^{-1-\bamma}\,\calu(z)\,,
\label{fromnutoz}
\end{eqnarray}
with 
\begin{eqnarray}
\tilde{\calu}(\nu,z_0)\equiv 
\int {d^2z_1 d^2z_2\over \pi^2z_{12}^4}\,\tilde{\calu}(z_1,z_2)\left(z_{12}^2\over z_{10}^2z_{20}^2\right)^\bamma\,.
\label{Unu3}
\end{eqnarray}
The symbol \textit{tilde} indicates that the Wilson lines run along $x^-$ direction as
\begin{eqnarray}
\tilde{U}(x_\perp) = {\rm P}{\rm exp}\left\{ig\int_{-\infty}^{+\infty} dx^-A^+(x^-,x_\perp)\right\}\,.
\end{eqnarray}
In eq. (\ref{IF}) we defined the two tensor structures $P_1^{\mu\nu} $ and $P_2^{\mu\nu} $ as
\begin{eqnarray}
P_1^{\mu\nu} = g^{\mu\nu} - {q^\mu q^\nu\over q^2}\,,
~~~~~~
P_2^{\mu\nu} = {1\over q^2}\left(q^\mu - {p_2^\mu q^2\over q\cdot p_2}\right)
\left(q^\nu - {p_2^\nu q^2\over q\cdot p_2}\right)\,.
\end{eqnarray}

\section{Correlation function with light-ray operators}

Correlation functions in Conformal Field Theory play an essential role: 
they encode information about the spectrum of the operators and their scaling dimensions. 
In two-dimensional CFTs, the primary fields and their descendants can be organized into representations of the Virasoro algebra,
and the correlation functions of these primary fields can be determined by conformal symmetry 
and the operator product expansion (OPE)~\cite{DiFrancesco:1997nk}.

In higher-dimensional CFTs, correlation functions can be analyzed using the conformal bootstrap method, which is based on the OPE and the crossing symmetry of the correlation functions. This approach has led significant progress in determining the scaling dimensions and OPE coefficients of higher-dimensional CFTs~\cite{Rychkov:2016iqz}.

In CFTs, local operators, the observables of the theory, are classified according to their scaling dimensions and their transformation 
properties under the conformal group, and can be organized into representations of the conformal group called conformal multiplets.

Correlation functions, which are the expectation values of products of local operators at distinct spacetime points, 
are important for the understanding of the dynamics and symmetries of the theory. 
In CFTs the conformal invariance imposes strong constraints on the form of the correlation functions, leading to a set of algebraic relations known as the conformal bootstrap equations~\cite{Belavin:1984vu}.

Conformal invariance completely fixes the form of the two- and three-point functions
up to constants, known as the operator product expansion (OPE) coefficients. These coefficients play an essential role in the structure of the theory and can be calculated using various techniques, such as the conformal bootstrap method and integrability~\cite{Rychkov:2016iqz}.

$\cal N$=4 super Yang-Mills theory is a four-dimensional quantum field theory with maximal supersymmetry, described by a gauge field, 
four Weyl fermions, and six scalar fields, all transforming in the adjoint representation of the gauge group~\cite{Green:1982sw}. 
The $\cal N$=4 SYM Lagrangian is invariant under supersymmetry transformations, which relate the bosonic and fermionic degrees of freedom. 
This invariance leads to a rich structure of non-renormalization theorems and exact results that make
$\cal N$=4 SYM particularly tractable~\cite{DHoker:1999yni}.

Similar to CFTs, correlation functions in N=4 SYM are central to the understanding the dynamics of the theory. 
However, the presence of supersymmetry imposes additional constraints on the correlation functions, simplifying their calculation.

In this section we are going to consider the corelation functions of
the supermultiplet of twist-two conformal operators, which is a fundamental concept in $\cal N$=4 super-Yang-Mills theory. 
A supermultiplet is a collection of states that transform under the same irreducible representation of the superconformal algebra. 
Twist-two operators are a class of operators with specific scaling dimensions and transformation properties under conformal transformations. 
They are important because they allow us to study the renormalization properties of the theory by revealing the structure of the theory under conformal transformations.

The twist-two supermultiplets can be classified according to their collinear twist. 
The term "collinear twist" refers to the difference between the scaling dimension and the collinear spin of the operator, 
which determines how the operator behaves under conformal transformations. 
Operators sharing the same collinear twist form a multiplet with the same properties. 
For example, they have the same scaling behavior under renormalization and similar transformation properties under conformal transformations.

In the context of the superconformal algebra, the irreducible representations are the fundamental building blocks that
describe how operators transform under the action of the superconformal algebra. Each irreducible representation is characterized by a set of quantum numbers, such as scaling dimension, spin, and R-charge, which determine the transformation properties of the operators in that representation. The classification of twist-two multiplets by collinear twist is essential for identifying the irreducible representations of the superconformal algebra. When studying the conformal structure of $\cal N$=4 super-Yang-Mills theory, 
it is important to understand how the operators transform under the action of the superconformal algebra. 
By classifying the twist-two multiplets according to their collinear twist, it becomes possible to identify the corresponding 
irreducible representations and to analyze their transformation properties in a systematic way~\cite{Belitsky:2003sh}.

\subsection{Twist-two non-local operator with non-integer spin $j$}

In this section it is convenient to introduce the light-cone vectors $p_1^\mu = \sqrt{s/2}\, n^\mu$ and $p_2^\mu = \sqrt{s/2}\,n'^\mu$ 
such that $p_1\cdot p_2 = s/2$. We will also use the notation 
$x_{p_1}= p_1^\mu x_\mu = \sqrt{s/2}\,x^-$, and $x_{p_2}= p_2^\mu x_\mu = \sqrt{s/2}\,x^+$.

To study the correlation function of operators with spin $ j$ and $j'$ in the high-energy limit, \textit{i.e.} in the limit in which
the contributions ${\alpha_s\over j-1}\sim 1$ are dominant and needs to be resummed with BFKL, we need to consider
the analytic continuation of local operators to non-integers values of spin $j$.
In supersymmetric $\cal N$=4 gauge theory one can construct the super-multiplet of non-local operators as an analytic continuation
of the local ones to non-integer values of spin $j$~\cite{Belitsky:2003sh, Balitsky:2018irv} (see also \cite{Chirilli:2021euj})
\begin{eqnarray}
&&\calf^j_{p_1}(x_\perp) = \int_0^{\infty}\! du\, u^{1-j}\calf_{p_1}(up_1+x_\perp)\,,
\label{calfspinj}\\
&&\Lambda^j_{p_1}(x_\perp) = \int_0^{\infty}\! du\, u^{-j}\Lambda_{p_1}(up_1+x_\perp)\,,
\label{Lambdaspinj}\\
&&\Phi^j_{p_1}(x_\perp) = \int_0^\infty\! du\, u^{-1-j}\Phi_{p_1}(up_1+x_\perp)\,,
\label{Phispinj}
\end{eqnarray}
with
\begin{eqnarray}
&&\calf_{p_1}(up_1,x_\perp) = \int \!dv 
\,{F^a}_{p_1\mu}(up_1+vp_1 + x_\perp)[up_1+vp_1,vp_1]_x^{ab}{{F^b}_{p_1}}^\mu(vp_1+x_\perp)\,,\\
&&\Lambda_{p_1}(up_1,x_\perp) = {i\over 2}\int \!dv\Big(-\bar{\lambda}^a_A(up_1+vp_1 + x_\perp)
[up_1+vp_1,vp_1]_x^{ab}\sigma_{p_1}\lambda^b_A(vp_1+x_\perp)
\nonumber\\
&&~~~~~~~~~~~~~~~~~~~~~ + \bar{\lambda}^a_A(vp_1 + x_\perp)
[vp_1, up_1+vp_1]_x^{ab}\sigma_{p_1}\lambda_A^b(up_1+vp_1+x_\perp)\Big)\,,\\
&&\Phi_{p_1}(u,x_\perp) = \int\!dv\,\phi_I^a(up_1+vp_1+x_\perp)[up_1+vp_1,vp_1]_x^{ab}\phi_I^b(vp_1+x_\perp) \,.
\end{eqnarray}
The corresponding multiplicatively renormalizable light-ray operators with non-integer $j$ for forward matrix elements are
\begin{eqnarray}
&&\cals_1 = \calf^j_{p_1} + {j-1\over 4}\Lambda^j_{p_1} - j(j-1)\half \Phi^j_{p_1}\,,
\label{S1Forwardj}
\\	
&&\cals_2 =  \calf^j_{p_1} - {1\over 4}\Lambda^j_{p_1} + {j(j+1)\over 6}\Phi^j_{p_1}\,,
\label{S2Forwardj}
\\
&&\cals_3 =  \calf^j_{p_1} - {j+2\over 2}\Lambda^j_{p_1} - {(j+1)(j+2)\over 2}\Phi^j_{p_1}\,.
\label{S3Forwardj}
\end{eqnarray}
As demonstrated in ref.~\cite{Balitsky:2014sqe}, the analytic continuation of anomalous dimensions of local operator to non integer $j$
gives the anomalous dimension of light-ray operators. So, the anomalous dimensions are
\begin{eqnarray}
\gamma_j^{\cals_1}(\alpha_s) = 4[\psi(j-1) + \gamma_E] + \calo(\alpha^2_s)\,, ~~~ \gamma_j^{\cals_2} = \gamma_{j+2}^{\cals_1}\,,
~~~ \gamma_j^{\cals_3} = \gamma^{\cals_1}_{j+4}\,.
\end{eqnarray}
In CFT the two-point correlators of light-ray operators is entirely fixed by conformal symmetry 
up to some unknown structure constant like for the correlators of local operators. So, we may write
\begin{eqnarray}
	\langle S^j(z_{1\perp})S^{j'}(z_{2\perp})\rangle = \delta(j-j'){C(j,\Delta)s^{j-1}\over[ (z_{1\perp}-z_{2\perp})^2]^{\Delta -1}}\mu^{-2\gamma_{\rm an}}
	\label{2point-general}
\end{eqnarray}
where in (\ref{2point-general}) $\Delta$ is the dimension of the operator and $\gamma_{\rm an}$ is the anomalous dimension.
\begin{figure}[t!]
	\begin{center}
		\includegraphics[width=5.0in]{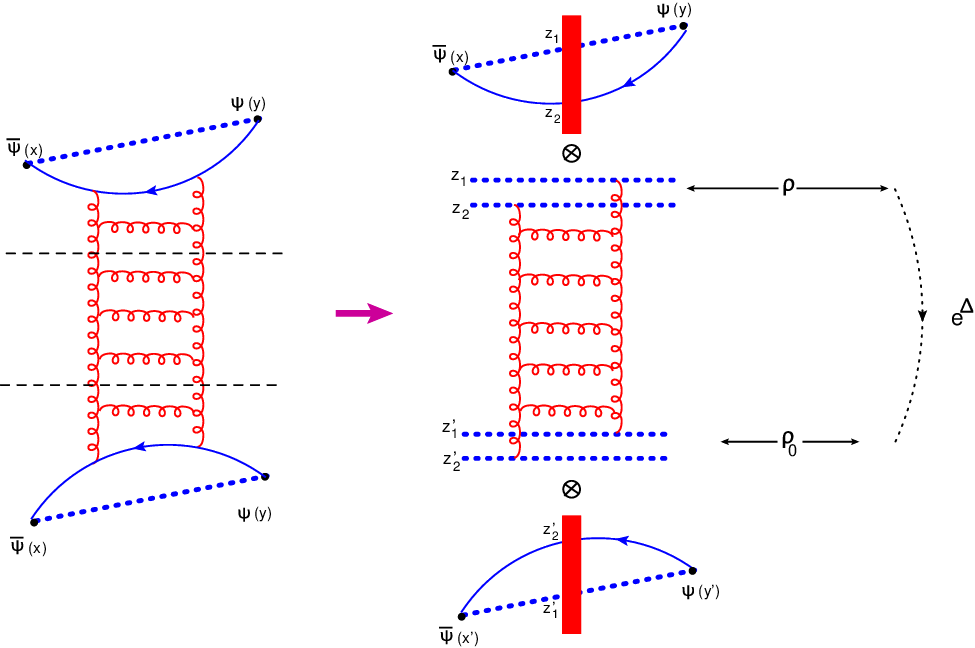}
		\caption{Here we present a diagrammatic representation of the four-point fermion correlation function. The left diagram illustrates the four-point function utilizing the BFKL approximation, whereas the right diagram demonstrates the four-point function with the application of the high-energy OPE.}
		\label{Fig:4correlation}
	\end{center}
\end{figure}

Our aim is to calculate the gluino correlation function (see Fig. \ref{Fig:4correlation})
\begin{eqnarray}
\langle\Lambda^j_{p_1}(x_\perp) \Lambda^j_{p_2}(x'_\perp)\rangle
\label{onlighcone}
\end{eqnarray}
in the BFKL limit. However, it is known that correlation functions of non-local operator on the light-cone are divergent in the BFKL limit
\textit{i.e.} in the limit of $j\to 1$ with ${\alpha_s\over j-1}\sim 1$. For this reason, to regulate the UV divergences 
one may adopt the point-splitting regulator, thus defying the set of point-splitting
super-multiplet of non-local operator with non-integer $j$ as
\begin{eqnarray}
&&\calf^j_{p_1}(x_\perp,y_\perp) = \int_0^{+\infty}\!du^{1-j} \calf_{p_1}(u;x_\perp,y_\perp)\,,\\
&&\Lambda^j_{p_1}(x_\perp,y_\perp) = \int_0^{+\infty}\!du^{1-j} \Lambda_{p_1}(u;x_\perp,y_\perp)\,,\\
&&\Phi^j_{p_1}(x_\perp,y_\perp) = \int_0^{+\infty}\!du^{1-j} \Lambda_{p_1}(u;x_\perp,y_\perp)\,,
\end{eqnarray}
with
\begin{eqnarray}
\hspace{-0.5cm}\calf_{p_1}(u;x_\perp,y_\perp) = \!\!&&\int d v {F^{a}}_{p_1 i}(up_1+ vp_1+x_\perp)
\nonumber\\
&&\times[up_1+vp_1+x_\perp, vp_1+y_\perp]^{ab}{{F^b}_{p_1}}^i(vp_1+y_\perp)\,,\\
\hspace{-0.5cm}\Lambda_{p_1}(u;x_\perp,y_\perp) = \!\!&&{i\over 2}\int \!dv\Big(-\bar{\lambda}^a_A(up_1+vp_1 + x_\perp)
[up_1+vp_1 +x_\perp ,vp_1+y_\perp]^{ab}\sigma_{p_1}\lambda^b_A(vp_1+y_\perp)
\nonumber\\
&& + \bar{\lambda}^a_A(vp_1 + x_\perp)
[vp_1 + x_\perp, up_1+vp_1+y_\perp]^{ab}\sigma_{p_1}\lambda_A^b(up_1+vp_1+y_\perp)\Big)\,,\\
\hspace{-0.5cm}\Phi_{p_1}(u;x_\perp,y_\perp) = \!\!&&\int\!dv\,\phi_I^a(up_1+vp_1+x_\perp)
[up_1+vp_1+x_\perp,vp_1+y_\perp]^{ab}\phi_I^b(vp_1+y_\perp) \,.
\label{quasipdf-def}
\end{eqnarray}

Let us rewrite the operators $\Lambda^j$ and $\calf^j$ in terms of operators $\cals_1, \cals_2, \cals_3$
using (\ref{S1Forwardj}), (\ref{S2Forwardj}), and (\ref{S3Forwardj})
\begin{eqnarray}
	&&\Lambda^j = {4(1+j)\over 3j^2}S_1^j - {4(1+2j)\over j^2(3+2j)}S_2^j + {4(1-2j)\over 3j(3+2j)}S_3^j
	\label{Lambdas123}
	\\
	&&\calf^j = {(1+j)(2+j)\over 6j^2}S_1^j - {(1-j)(2+j)(1+3j)\over 2j^2(3+2j)}S_2^j + {(1-j)(2-j)\over 6j(3+2j)}S_3^j
	\label{Fs123}
\end{eqnarray}
Using (\ref{Fs123}) the correlator with $\calf^j$ operators can be written in terms of the
$\cals$'s operators as
\begin{eqnarray}
	&&\hspace{-0.5cm}\left\langle\calf^j_{n_1}\left(x_\perp, y_\perp\right)
	\calf^{j'}_{n_2}\left(x'_\perp, y'_\perp\right)\right\rangle
	\label{gluoncorre-nonlcalop-def}\\
	\nonumber\\
	&&\hspace{-0.5cm}
	\stackrel{\Delta_\perp,\Delta'_\perp\to 0}{=} \delta(j-j') \Bigg[{(1+j)^2(2+j)^2\over 36j^4}{C_1(j,\Delta(j))s^{j-1}\over [(X_\perp-X'_\perp)^2]^{\Delta(j)-1}}\,
	|\Delta_\perp\Delta'_\perp|^{-2\gamma_j^{S_1}}
	\nonumber\\
	&&\hspace{2.7cm}
	+  {(1-j)^2(2-j)^2(1+3j)^2\over 4j^4(3+2j)^2}{C_2(j,\Delta(j))s^{j-1}\over [(X_\perp-X'_\perp)^2]^{\Delta(j)-1}}\,
	|\Delta_\perp\Delta'_\perp|^{-2\gamma_{j+2}^{S_1}}
	\nonumber\\
	&&\hspace{2.7cm}
	+  {(1-j)^2(2-j)^2\over 36j^2(3+2j)^2}{C_3(j,\Delta(j))s^{j-1}\over [(X_\perp-X'_\perp)^2]^{\Delta(j)-1}}\,
	|\Delta_\perp\Delta'_\perp|^{-2\gamma_{j+4}^{S_1}}\Bigg]
	\nonumber
\end{eqnarray}
with $X_\perp = {x_\perp+y_\perp\over 2}$, $X'_\perp = {x'_\perp+y'_\perp\over 2}$,
and $\Delta_\perp = x_\perp-y_\perp$ and the same for $\Delta'_\perp$.

Comparing (\ref{gluoncorre-nonlcalop-def}) with (\ref{2point-general}) we can appreciate how the point-splitting
act as an UV regulator. Indeed, the UV regulator $\mu$ in (\ref{gluoncorre-nonlcalop-def}) is now given by
$|\Delta_\perp\Delta'_\perp|$: if the transverse distances go to zero we get UV divergence.

We now observe that, taking the limit $j\to 1$, which is the high-energy limit as we discussed above,
the correlation function with $\calf^j$ operators is given only in terms of the $\cals_1$ operator so we have
\begin{eqnarray}
	&&\hspace{-0.5cm}\left\langle\calf^j_{n_1}\left(x_\perp, y_\perp\right)
	\calf^{j'}_{n_2}\left(x'_\perp, y'_\perp\right)\right\rangle
	\nonumber\\
	\nonumber\\
	&&\hspace{-0.5cm}
	\stackrel{\Delta_\perp,\Delta'_\perp\to 0}{=} \delta(j-j') {(1+j)^2(2+j)^2\over 36j^4}{C_1(j,\Delta(j))s^{j-1}\over [(X_\perp-X'_\perp)^2]^{\Delta(j)-1}}\,
	|\Delta_\perp\Delta'_\perp|^{-2\gamma_j^{S_1}}
	\label{gluoncorre-nonlcalop-def2}
\end{eqnarray}
Now, le us consider the correlation function for gluino operator (\ref{Lambdas123})
\begin{eqnarray}
	&&\hspace{-0.5cm}\left\langle\Lambda^j_{n_1}\left(x_\perp,y_\perp\right)
	\Lambda^{j'}_{n_2}\left(x'_\perp,y'_\perp\right)\right\rangle
	\nonumber\\
	&&\hspace{-0.5cm}
	\stackrel{\Delta_\perp,\Delta'_\perp\to 0}{=} \delta(j-j') {16(1+j)^2\over 9j^2}{C_1(j,\Delta)s^{j-1}\over  [(X_\perp-X'_\perp)^2]^{\Delta(j)-1}}\,
	|\Delta_\perp\Delta'_\perp|^{-2\gamma_j^{S_1}}
	\nonumber\\
	&&\hspace{-0.5cm}
	+ \delta(j-j') {16(1+2j)^2\over j^2(3+2j)^2}{C_2(j,\Delta)s^{j-1}\over [(X_\perp-X'_\perp)^2]^{\Delta(j)-1}}\,
	|\Delta_\perp\Delta'_\perp|^{-2\gamma_{j+2}^{S_1}}
	\nonumber\\
	&&\hspace{-0.5cm}
	+ \delta(j-j') {16(1-2j)^2\over 9j^2(3+2j)^2}{C_3(j,\Delta)s^{j-1}\over  [(X_\perp-X'_\perp)^2]^{\Delta(j)-1}}\,
	|\Delta_\perp\Delta'_\perp|^{-2\gamma_{j+4}^{S_1}}
	\label{gluinocorre-nonlcalop-def}
\end{eqnarray}

Therefore, the  correlation function with gluinos we have to consider is
\begin{eqnarray}
\langle\Lambda_{p_1}(x_\perp,y_\perp)\Lambda_{p_2}(x'_\perp, y'_\perp)\rangle
\label{gluinocorrefunct}
\end{eqnarray}
and study it at high-energy (Regge limit) $x^+, x'^+\to\infty$ and $y^-,y'^-\to -\infty$ keeping all other components fixed. 
In this limit the correlation function (\ref{gluinocorrefunct}) factorizes as~\cite{Balitsky:2009yp} (see Fig. \ref{Fig:4correlation})
\begin{eqnarray}
\langle\Lambda_{p_1}(x_\perp,y_\perp)\rangle\langle\Lambda_{p_2}(x'_\perp, y'_\perp)\rangle\,.
\label{facto-gluinocorrefunct}
\end{eqnarray}
Before proceeding with the calculation, we rewrite the operator $\Lambda^j_{p_1}$ as 
\begin{eqnarray}
\Lambda_{p_1}^j(x_\perp,y_\perp) =\!\!&& 
\left(s\over 2\right)^{{j-1\over 2}}\!\!\int_0^{+\infty}\!\!dL\,L^{-j}
{i\over 2}\!\!\int \!\! dx^+\Big(-\bar{\lambda}^a_A(x^+n+y^+n+x_\perp)
\label{energyscale}\\
&&\times
[x^+n+x^+n+x_\perp,y^+n+y_\perp]^{ab}\sigma^+\lambda^b_A(y^+n+y_\perp)
\nonumber\\
&&+\bar{\lambda}^a_A(y^+n+x_\perp)[y^+n + x_\perp, x^+n+y^+n+y_\perp]^{ab}
\sigma^+\lambda^b_A(x^+n+y^+n+y_\perp)\Big)\nonumber
\end{eqnarray}

Now, each factor of (\ref{facto-gluinocorrefunct}) we can apply the HE-OP which is diagrammatically shown in Fig.
and arrive at 
\begin{eqnarray}
&&\hspace{-0.5cm}\int_0^{+\infty} dx^+ \int_{-\infty}^0dy^+\delta(x^+-y^+-L)
\langle i\,\bar{\lambda}(x^+,x_\perp)\gamma^-[x^+n^\mu+x_\perp, y^+n^\mu+y_\perp]\lambda(y^+,y_\perp)\rangle_{{\rm Fig.}\ref{Fig:quarkLOif}}
\nonumber\\
&&\hspace{-0.5cm}
\times\! \int_0^{+\infty} dx'^- \int_{-\infty}^0dy'^-\delta(x'^--y'^--L')
\langle i\,\bar{\lambda}(x'^-,x'_\perp)\gamma^+[x'^-n'^\mu+x'_\perp, y'^-n'^\mu+y'_\perp]\lambda(y'^-,y'_\perp)\rangle_{{\rm Fig.}\ref{Fig:quarkLOif}}
\nonumber\\
&&\hspace{-0.5cm}
= \int_0^{+\infty} dx^+ \int_{-\infty}^0dy^+\delta(x^+-y^+-L)
\int_0^{+\infty} dy'^- \int_{-\infty}^0dx'^-\delta(x'^--y'^--L')
\nonumber\\
&&\hspace{-0.5cm}
~~\times{- 1\over x^{+2}y^{+2}}\!\!\int {d^2z_2\over \pi^3}
{\Big[(x-z_2)^2_\perp+(y-z_2)^2_\perp - (x-y)^2_\perp\Big]\over \Big[{(y-z_2)^2_\perp\over |y^+|} + 
{(x-z_2)^2_\perp\over x^+}+i\epsilon\Big]^3}\,\big[-2N^2_c\calu(z_1,z_2)\big]
\nonumber\\
&&\hspace{-0.5cm}
~~\times{- 1\over x'^{-2}y'^{-2}} \!\!\int {d^2z'_2\over \pi^3}
{\Big[(x'-z'_2)^2_\perp+(y'-z'_2)^2_\perp - (x'-y')^2_\perp\Big]\over \Big[{(y'-z'_2)^2_\perp\over |y'^-|} 
+ {(x'-z'_2)^2_\perp\over x'^-}+i\epsilon\Big]^3}\,\big[-2N^2_c\tilde{\calu}(z'_1,z'_2)\big]
\label{gluinogluino1}
\end{eqnarray}
where we remind that $z_1 = ux_\perp + \baru y_\perp$ and $u={|y^+|\over \Delta^+}$ and $\baru = {x^+\over \Delta^+}$,
and $\Delta^+=x^++|y^+| = L$. In eq. (\ref{gluinogluino1}) we introduced the operator
$\tilde{\calu}_{z_1z_2} = 1 - {1\over N_c}\tr\{\tilde{U}_{z_1}\tilde{U}^\dagger_{z_2}\}$ with the Wilson line
along $n'^\mu$ direction defined as
\begin{eqnarray}
\tilde{U}_x = \tilde{U}(x_\perp) = {\rm P}{\rm exp}\left\{ig\int_{-\infty}^{+\infty} dx^-\,A^+(x^-n'+x_\perp) \right\}\,.
\end{eqnarray}
Moreover, since the gluino lives in the adjoint representation, we also need the substitution 
$U_{z_2}^{ab}U_{z_1}^{ab} \to -2N^2_c\calu(z_{12})$.
To complete the calculation of the correlation function at high-energy we need the scattering of the two dipole-Wilson lines
at the leading-log approximation. To this end we use the solution of the evolution equation for the dipole-Wilson line
operator in the linear case, \textit{i.e.} the BFKL equation,
\begin{eqnarray}
&&\calu^{Y_a}(\nu,z_0) = e^{(Y_a-Y_0)\aleph(\nu)}\calu^{Y_0}(\nu,z_0)\,,
\label{calu-nua}\\
&&\calu^{Y_b}(\nu,z'_0) = e^{(Y_0+Y_b)\aleph(\nu)}\calu^{Y_0}(\nu,z'_0)\,,
\label{calu-nub}
\end{eqnarray}
where $\aleph(\nu) = \aleph(\gamma(\nu))$ with $\gamma = \half + i\nu$.
The operator $\calu(\nu,z_0)$ is defined through the projection onto the LO eigenfunctions as
\begin{eqnarray}
\calu(\nu,z_0) \equiv \int {d^2 z'_1 d^2z'_2\over\pi^2z'^4_{12}}
\left({z'^2_{12}\over z'^2_{10}z'^2_{20}}\right)^\bamma\calu(z'_1,z'_2)\,,
\label{calu-nu1}
\end{eqnarray}
and using the completeness relation of the LO eigenfunctions we have
\begin{eqnarray}
&&\hspace{-2cm}\calu(z_1,z_2)= 
\int\!d^2z_0 \int {d\nu\over \pi^2}\,\nu^2\left({z^2_{12}\over z^2_{10}z^2_{20}}\right)^\gamma \calu(\nu,z_0)\,,
\label{calu-nu}
\end{eqnarray}
with $\gamma = \half + i\nu$ and $\bamma = 1-\gamma$.
Implementing eqs. (\ref{calu-nua}), (\ref{calu-nub}), (\ref{calu-nu1}), and (\ref{calu-nu})
we obtain the dipole-dipole scattering in the LLA as (see Ref.~\cite{Balitsky:2009yp} and Appendix C of Ref.~\cite{Chirilli:2021euj} 
for further details on the dipole-dipole scattering calculation)
\begin{eqnarray}
&&\hspace{-1cm}
\langle \calu^{Y_a}(z_1,z_2)\tilde{\calu}^{Y_b}(z'_1,z'_2)\rangle
\nonumber\\
&&\hspace{-1cm}
= - {\alpha^2_s (N^2_c-1)\over N^2_c}\int {d\nu \over \pi}{16\,\nu^2\over (1+4\nu^2)^2}
\Big({2LL'\over |(x-y)_\perp||(x'-y')_\perp|}\Big)^{\aleph(\gamma)}
\nonumber\\
&&\hspace{-1cm}
~~~\times{\Gamma^2(\half + i\nu)\Gamma(-2i\nu)\over\Gamma^2(\half - i\nu)\Gamma(1+2i\nu)} 
\left({z^2_{12}z'^2_{12}\over (X-X')^4}\right)^\gamma\,,
\label{dipo-dipo}
\end{eqnarray}
where, as done above, we defined $X_\perp = {x_\perp+y_\perp\over 2}$ and the same for $X'_\perp$ and we used 
\begin{eqnarray}
e^{(Y_a+Y_b)\aleph(\nu)} = \left(2L L'\over |(x-y)_\perp||(x'-y')_\perp|\right)^{\aleph(\gamma)}\,,
\end{eqnarray}
with resummation parameter the rapidities~\cite{Balitsky:2009yp}
\begin{eqnarray}
&& Y_a = \half\ln{2L^2\over (x-y)^2_\perp}\,, ~~~~~~~~ Y_b = \half\ln{2L'^2\over (x'-y')^2_\perp}\,.
\label{resumparameter-coord}
\end{eqnarray}
Using result (\ref{dipo-dipo}) in (\ref{gluinogluino1}) 
for operator $\Lambda_{p_1}$ is
\begin{eqnarray}
&&\hspace{-1cm}\langle\Lambda_{p_1}(x_\perp,y_\perp)\Lambda_{p_2}(x'_\perp, y'_\perp)\rangle
\nonumber\\
&&\hspace{-1cm} = -{32\alpha^2_s\over \pi^4}\left(s\over 2\right)^{j-1} {N^2_c(N^2_c-1)\over \Delta^2_\perp\Delta'^2_\perp}\!\int {d\nu\over \pi} 
{8\,\nu^2\over (1+4\nu^2)^2}
{\left(2L L'\right)^{\aleph(\gamma)}\over [(X-X')_\perp^2]^{2\gamma}}
\nonumber\\
&&\hspace{-1cm}
~~~\times{\gamma^4\Gamma^8(\gamma)\Gamma(1-2\gamma)\over \Gamma^3(2\gamma)}
[\Delta^2_\perp\Delta'^2_\perp]^{\gamma - {\aleph(\gamma)\over 2}}
\label{q-correlation-step1}
\end{eqnarray}
 where we used
\begin{eqnarray}
&&\hspace{-2cm}i\int_0^{+\infty}\!\! dx^+ \!\!\int_{-\infty}^0dy^+\delta(x^+-y^+-L)
\nonumber\\
&&\hspace{-2cm}
\times\!\int {d^2z_2\over x^{+2}y^{+2}}
{\Big[(x-z_2)^2_\perp+(y-z_2)^2_\perp - (x-y)^2_\perp\Big]\over 
\Big[{(y-z_2)^2_\perp\over |y^+|} + {(x-z_2)^2_\perp\over x^+}+i\epsilon\Big]^3} z^{2\gamma}_{12}
= {4\,i\,\pi^2 \over [\Delta^2_\perp]^{1-\gamma}}
{\gamma^2B(\gamma)\over \sin(\pi\gamma)}
\label{project-longint}
\end{eqnarray}
and the same result with replacement $\Delta_\perp\to \Delta'_\perp$ for the integration over $x'^+$ and $y'^+$
for the impact factor with operators along the light-cone vector $n'$. In eq. (\ref{project-longint})
we used $B(\gamma) = {\Gamma^2(\gamma)\over \Gamma(2\gamma)}$.

We need the correlation function for the 
of the gluino operators with specific spins $j$ and $j'$, so using
eq. (\ref{Lambdaspinj}) we have
\begin{eqnarray}
&&\hspace{-0.5cm}\int_0^{+\infty}dL\,L^{-j}\int _0^{+\infty} dL'\,L'^{-j'}
\int_0^{+\infty} dx^+ \int_{-\infty}^0dy^+\delta(x^+-y^+-L)
\nonumber\\
&&\hspace{-0.5cm}
\times\!\! \int_0^{+\infty} dx'^- \int_{-\infty}^0dy'^-\delta(x'^--y'^--L')
\langle \,i\bar{\lambda}(x^+,x_\perp)\gamma^-[x^+n^\mu+x_\perp, y^+n^\mu+y_\perp]\lambda(y^+,y_\perp)\rangle
\nonumber\\
&&\hspace{-0.5cm}
\times\langle\,i\bar{\lambda}(x'^-,x'_\perp)\gamma^+[x'^-n'^\mu+x'_\perp, y'^-n'^\mu+y'_\perp]\lambda(y'^-,y'_\perp)\rangle
\theta(2LL'-(X-X')^2_\perp)
\nonumber\\
&&\hspace{-0.5cm}
= {i8\alpha^2_s\over \pi^4}N^2_c (N^2_c-1)
\int_{\half -i\infty}^{\half +i\infty} d\gamma \,{(1-2\gamma)^2\over (1-\gamma)^2}
{2^\omega\,[\Delta^2_\perp\Delta'^2_\perp]^{\gamma - 1}\over [(X-X')^2_\perp]^{2\gamma+\omega}}
{\gamma^2\Gamma^8(\gamma)\Gamma(1-2\gamma)\over \Gamma^3(2\gamma)}
\nonumber\\
&&\hspace{-0.5cm}
~~~\times{\theta\big({\rm Re}[j-1-\aleph(\gamma)]\big)\over \aleph(\gamma) - \omega }
\left((X-X')^2_\perp \over |\Delta_\perp||\Delta'_\perp|\right)^{\aleph(\gamma)}\delta(\varsigma'-\varsigma)
\label{q-correlation-step2}
\end{eqnarray}
with $\omega=j-1$ and where we used 
\begin{eqnarray}
&&\hspace{-2cm}
\int_0^{+\infty}\!dL\, L^{-j}\int_0^{+\infty}\!dL'\, L'^{-j'}(LL')^{\aleph(\gamma)}
\,\theta\left(2LL'-(X-X')^2_\perp\right) 
\nonumber\\
&&\hspace{-2cm}
=  {\theta\big({\rm Re}[j-1-\aleph(\gamma)]\big)\over j-1-\aleph(\gamma)}
\left[2^{-1}(X-X')^2_\perp \right]^{1-j+\aleph(\gamma)}(2\pi)\delta(\varsigma'-\varsigma)
\label{doublemellin}
\end{eqnarray}
with $j = a + i\varsigma$, $j' = a + i\varsigma'$, $a\in \Re$, and $a>1+\aleph(\gamma)$.
The next step is to perform the integration over $\gamma$. To this end, 
we recall that we are in the limit of small $\Delta^2_\perp$ and $\Delta'^2_\perp$,
so we can calculate the integral by taking the residues closing the contour to the right. Therefore, we need to consider the
residue at $\aleph(\gamma)-\omega=0$ and at $\gamma=1$.

Let $\tilde{\gamma}$ be the solution of $\aleph(\tilde{\gamma})-\omega=0$ and expanding around $\tilde{\gamma}$ we have
$\aleph(\tilde{\gamma}) + (\gamma-\tilde{\gamma})\aleph'(\tilde{\gamma})$. The residue at $\tilde{\gamma}$ is
\begin{eqnarray}
&&\hspace{-0.3cm}\int_0^{+\infty}dL\,L^{-j}\int _0^{+\infty} dL'\,L'^{-j'}
\int_0^{+\infty} dx^+ \int_{-\infty}^0dy^+\delta(x^+-y^+-L)
\nonumber\\
&&\hspace{-0.3cm}
\times\!\! \int_0^{+\infty} dx'^- \int_{-\infty}^0dy'^-\delta(x'^--y'^--L')
\langle \,i\bar{\lambda}(x^+,x_\perp)\gamma^-[x^+n^\mu+x_\perp, y^+n^\mu+y_\perp]\lambda(y^+,y_\perp)\rangle
\nonumber\\
&&\hspace{-0.3cm}
\times\langle\,i\bar{\lambda}(x'^-,x'_\perp)\gamma^+[x'^-n'^\mu+x'_\perp, y'^-n'^\mu+y'_\perp]\lambda(y'^-,y'_\perp)\rangle
\theta(2LL'-(X-X')^2_\perp)
\nonumber\\
&&\hspace{-0.3cm}
= \alpha^2_s N^2_c(N^2_c-1){16\over \pi^3}
{(1-2\gamma^*)^2\over (1-\gamma^*)^2}
{{\gamma^*}^2\Gamma^8(\gamma^*)\Gamma(1-2\gamma^*)\over N'(\gamma^*)\Gamma^3(2\gamma^*)}
{2^\omega|\Delta_\perp\Delta'_\perp|^{2\gamma^* - 2 - \omega}\over [(X-X')^2_\perp]^{2\gamma^*}}
\delta(\varsigma'-\varsigma)
\label{quark-correlation}
\end{eqnarray}
In result (\ref{quark-correlation}), $\Delta_\perp\Delta'_\perp$ is like the UV cut-off $\mu^{-2}$ so comparing with 
the general expression for the two-point function in CFT, (\ref{2point-general}), we
obtain an equation which relate the  equation $2\gamma^* - 2 - \omega = \gamma_{\rm an}$
and $2\gamma^* = \Delta - 1$ so we get that $\gamma_{\rm an} + \omega = \Delta - 3$.
From this we conclude that the anomalous dimension of light-ray operators 
in the BFKL limit is given by the solution of equation $\omega =\aleph(\Delta)$.

Let us now consider the overlapping region between the BFKL and the DGLAP regime, that is
we now consider $\alpha_s\ll\omega\ll 1$. So, using (\ref{alephexpand}),
for $\gamma\to 1$ we have $\aleph(\gamma)\to{\bar{\alpha}_s\over 1-\gamma}$ and
${1\over \omega-\aleph(\gamma)}\to {1\over \omega - {\bar{\alpha}_s\over 1-\gamma}}=
{1-\gamma\over -\omega(\gamma-1+{\bar{\alpha}_s\over \omega})}$. Using this in (\ref{q-correlation-step2}),
we have
\begin{eqnarray}
&&\hspace{-1.5cm}
\int_0^{+\infty}dL\,L^{-j}\int_0^{+\infty} dL'\,L'^{-j'}
\int_0^{+\infty} dx^+ \int_{-\infty}^0dy^+\int_0^{+\infty} dx'^- \int_{-\infty}^0dy'^-
\nonumber\\
&&\hspace{-1.5cm}
\times\! \langle i\bar{\lambda}(x^+,x_\perp)\gamma^-[x^+n^\mu+x_\perp, y^+n^\mu+y_\perp]\lambda(y^+,y_\perp)\rangle
\delta(x^+-y^+-L)
\nonumber\\
&&\hspace{-1.5cm}
\times\! 
\langle i\bar{\lambda}(x'^-,x'_\perp)\gamma^+[x'^-n'^\mu+x'_\perp, y'^-n'^\mu+y'_\perp]\lambda(y'^-,y'_\perp)\rangle
\delta(x'^--y'^--L')
\nonumber\\
&&\hspace{-1.5cm}
=  i\alpha^2_s N^2_c(N^2_c-1){8\over \pi^4}
\int_{\half -i\infty}^{\half +i\infty} d\gamma \,{(1-2\gamma)^2\over (1-\gamma)}
{2^\omega\,[\Delta^2_\perp\Delta'^2_\perp]^{\gamma - 1}\over [(X-X')^2_\perp]^{2\gamma+\omega}}
{\gamma^2\Gamma^8(\gamma)\Gamma(1-2\gamma)\over \Gamma^3(2\gamma)}
\nonumber\\
&&\hspace{-1.5cm}
~~\times\!
\left((X-X')^2_\perp \over |\Delta_\perp||\Delta'_\perp|\right)^{\bar{\alpha}_s\over 1-\gamma}
{\delta(\varsigma'-\varsigma)\over \omega(\gamma - 1 + {\bar{\alpha}_s\over \omega})}
\end{eqnarray}
Taking the residue at $\gamma = 1-{\alpha_s\over \omega}$ we finally have
\begin{eqnarray}
&&\hspace{-1.5cm}
\int_0^{+\infty}dL\,L^{-j}\int_0^{+\infty} dL'\,L'^{-j'}
\int_0^{+\infty} dx^+ \int_{-\infty}^0dy^+\int_0^{+\infty} dx'^- \int_{-\infty}^0dy'^-
\nonumber\\
&&\hspace{-1.5cm}
\times\! \langle i\bar{\lambda}(x^+,x_\perp)\gamma^-[x^+n^\mu+x_\perp, y^+n^\mu+y_\perp]\lambda(y^+,y_\perp)\rangle
\delta(x^+-y^+-L)
\nonumber\\
&&\hspace{-1.5cm}
\times\! 
\langle  i\bar{\lambda}(x'^-,x'_\perp)\gamma^+[x'^-n'^\mu+x'_\perp, y'^-n'^\mu+y'_\perp]\lambda(y'^-,y'_\perp)\rangle
\delta(x'^--y'^--L')
\nonumber\\
&&\hspace{-1.5cm}
=  - {8N^2_c(N^2_c-1)\over N^2_c\pi}
{2^\omega\,\omega\over [(X-X')^2_\perp]^{2-2{\bar{\alpha}_s\over \omega}}}
\left(1\over |\Delta_\perp||\Delta'_\perp|\right)^{\omega + 2{\bar{\alpha}_s\over \omega}}\delta(\varsigma'-\varsigma)
\label{gluinocorrelatorLT}
\end{eqnarray}
where we used 
\begin{eqnarray}
{(1-2\gamma)^2\gamma^2\Gamma^8(\gamma)\Gamma(1-2\gamma)
\over (1-\gamma)\Gamma^3(2\gamma)\omega(\gamma - 1 + {\bar{\alpha}_s\over \omega})}
\stackrel{\gamma=1-{\bar{\alpha}_s\over \omega}}{\simeq} - {\omega^2\over 2\bar{\alpha}_s^2} + \calo(\alpha_s^{-1})\,.
\end{eqnarray}
We rewrite result (\ref{gluinocorrelatorLT}) in terms of the operator $\Lambda^j_{p_1}$, and $\lambda^{j'}_{p_2}$,
and we have
\begin{eqnarray}
&&\hspace{-1.5cm}
\langle \Lambda^j_{p_1}(x_\perp, y_\perp)\Lambda^{j'}_{p_2}(x'_\perp, y'_\perp)\rangle
\simeq  - {8N^2_c\over \pi}
{\omega \,s^\omega \left( |\Delta_\perp||\Delta'_\perp|\right)^{-\omega - 2{\bar{\alpha}_s\over \omega}}\over [(X-X')^2_\perp]^{2-2{\bar{\alpha}_s\over \omega}}}
\delta(\varsigma'-\varsigma)
\label{gluinocorrelatorLT2}
\end{eqnarray}
where we used $N^2-1\simeq N^2_c$ at large $N_c$.

From eq. (\ref{gluinocorrelatorLT2}) we obtain that the anomalous dimension in the BFKL limit is 
$\gamma_{an}\simeq {\bar{\alpha}_s\over \omega}$.

From eq. (\ref{q-correlation-step1}), we notice that besides the moving (dynamical) poles, like the one we just calculated, there 
ia also residue at $\gamma=1$. To calculate its contribution, we start indeed from eq. (\ref{q-correlation-step1}), and expand near $\gamma=1$
obtaining
\begin{eqnarray}
&&\hspace{-0.5cm}\int_0^{+\infty}dL\,L^{-j}\int _0^{+\infty} dL'\,L'^{-j'}
\int_0^{+\infty} dx^+ \int_{-\infty}^0dy^+\delta(x^+-y^+-L)
\nonumber\\
&&\hspace{-0.5cm}
\times\!\! \int_0^{+\infty} dx'^- \int_{-\infty}^0dy'^-\delta(x'^--y'^--L')
\langle \,i\bar{\lambda}(x^+,x_\perp)\gamma^-[x^+n^\mu+x_\perp, y^+n^\mu+y_\perp]\lambda(y^+,y_\perp)\rangle
\nonumber\\
&&\hspace{-0.5cm}
\times\langle\,i\bar{\lambda}(x'^-,x'_\perp)\gamma^+[x'^-n'^\mu+x'_\perp, y'^-n'^\mu+y'_\perp]\lambda(y'^-,y'_\perp)\rangle
\theta(2LL'-(X-X')^2_\perp)
\nonumber\\
&&\hspace{-0.5cm}	= i\alpha^2_s N^2_c(N^2_c-1){8\over \pi^4}
\int_{\half -i\infty}^{\half +i\infty} \!d\gamma \, {2^\omega\over[(X-X')^2_\perp]^{2+\omega}}
\left({[(X-X')^2_\perp]^2 \over \Delta^2_\perp\Delta'^2_\perp}\right)^{1-\gamma}
\nonumber\\
&&\hspace{-0.5cm}
~~~\times{1\over \omega}{(1-2\gamma)^2\gamma^2\Gamma^8(\gamma)\Gamma(1-2\gamma)
\over (1-\gamma)\Gamma^3(2\gamma)(\gamma - 1 + {\bar{\alpha}_s\over \omega})}
\left((X-X')^2_\perp \over |\Delta_\perp||\Delta'_\perp|\right)^{\bar{\alpha}_s\over 1-\gamma}\delta(\varsigma'-\varsigma)
\end{eqnarray}
Notice that we now have a simple pole and an essential pole both at $\gamma=1$. 
Calculating the residue, keeping in mind that we are in the $\alpha_s\ll\omega$ limit, we have
\begin{eqnarray}
\langle \Lambda^j_{p_1}(x_\perp, y_\perp)\Lambda^{j'}_{p_2}(x'_\perp, y'_\perp)\rangle
\simeq&&  {8N^2_c\over\pi}
{\omega \, s^\omega\over[(X-X')^2_\perp]^{2+\omega}}
\nonumber\\
&&\times\Big[1 + 2{\bar{\alpha}_s\over \omega}\ln\left((X-X')^2_\perp \over |\Delta_\perp||\Delta'_\perp|\right)\Big]
	\delta(\varsigma'-\varsigma)
\end{eqnarray}
where we used $N^2-1\simeq N^2_c$ at large $N_c$.

The occurrence of such spurious poles would compromise the anticipated result from conformal symmetry, as presented in eq. (\ref{2point-general}). In my previous work~\cite{Chirilli:2021euj} and in ref.~\cite{Balitsky:2018irv}, 
it was illustrated that this contribution is offset by the diagrams not encompassed in the HE-OPE formalism, thus validating the expected conformal result (\ref{2point-general}). 
In a similar manner, here, these spurious poles will also be counteracted 
by contributions from other diagrams not accounted for within the HE-OPE formalism. 
At present, the mechanism enabling the cancellation of such spurious poles has yet to be discovered.
\bibliographystyle{JHEP}
\bibliography{/Users/chirilli/Documents/mm/m/MyReferences}

\end{document}